\documentclass{emulateapj}

\newcommand\asca{{\it ASCA}}

\newcommand\sax{{\it BeppoSAX}}
\newcommand\chandra{{\it Chandra}}

\newcommand\rosat{{\it ROSAT}}
\newcommand\rxte{{\it RXTE}}
\newcommand\xmm{{\it XMM-Newton}}

\newcommand\s{{\rm~s}}
\newcommand\ks{{\rm~ks}}
\newcommand\kev{{\rm~keV}}
\newcommand\ev{{\rm~eV}}
\newcommand\kms{\ifmmode {\rm~km\ s}^{-1} \else ~km s$^{-1}$\fi}
\newcommand\Hunit{\ifmmode {\rm~km\ s}^{-1}\ {\rm Mpc}^{-1}
        \else ~km s$^{-1}$ Mpc$^{-1}$\fi}
\newcommand\ctssec{\ifmmode {\rm~count\ s}^{-1} \else ~count s$^{-1}$\fi}
\newcommand\ergsec{\ifmmode {\rm~erg\ s}^{-1} \else
        ~erg s$^{-1}$\fi}
\newcommand\funit{\ifmmode {\rm~erg\ s}^{-1}\;{\rm cm}^{-2} \else
        ~ergs s$^{-1}$ cm$^{-2}$\fi}
\newcommand\phflux{\ifmmode {\rm~photon\ s}^{-1}\;{\rm cm}^{-2}
        \else   ~photon s$^{-1}$ cm$^{-2}$\fi}
\newcommand\efluxA{\ifmmode {\rm~erg\ s}^{-1}\;{\rm cm}^{-2}\;{\rm
        \AA}^{-1} \else ~erg s$^{-1}$ cm$^{-2}$ \AA$^{-1}$\fi}
\newcommand\efluxHz{\ifmmode {\rm~erg\ s}^{-1}\;{\rm cm}^{-2}\;{\rm
        Hz}^{-1} \else ~erg s$^{-1}$ cm$^{-2}$ Hz$^{-1}$\fi}
\newcommand\cc{\ifmmode {\rm~cm}^{-3} \else cm$^{-3}$\fi}
\newcommand\FWHM{\ifmmode {\rm~FWHM} \else ${\rm~FWHM}$\fi}
\newcommand\Msun{\ifmmode M_{\odot} \else $M_{\odot}$\fi}
\newcommand\Lsun{\ifmmode L_{\odot} \else $L_{\odot}$\fi}
\newcommand\ltsim{\raisebox{-.5ex}{$\;\stackrel{<}{\sim}\;$}}
\newcommand\gtsim{\raisebox{-.5ex}{$\;\stackrel{>}{\sim}\;$}}
\newcommand\hbeta{\ifmmode {\rm H}\beta \else H$\beta$\fi}
\newcommand\Kalpha{\ifmmode {\rm K}\alpha \else K$\alpha$\fi}
\newcommand\nh{\ifmmode N_{\rm H} \else N$_{\rm H}$\fi}

\begin{document}

\title{\xmm{} view of the ultra-luminous X-ray sources in M~51}

\author{Gulab C. Dewangan\altaffilmark{1}, Richard E.
  Griffiths\altaffilmark{1}, Manojendu Choudhury\altaffilmark{2},
  Takamitsu Miyaji\altaffilmark{1}, \& Nicholas J.
  Schurch\altaffilmark{1}} \altaffiltext{1}{Department of Physics,
  Carnegie Mellon University, 5000 Forbes Ave, Pittsburgh, PA 15213
  USA} \altaffiltext{2}{National Center for Radio Astrophysics, Tata
  Institute of Fundamental Research, Pune 411007, India}

\begin{abstract}
  We present results based on \xmm{} observation of the nearby spiral
  galaxy M~51 (NGC~5194 and NGC~5195).  We confirm the presence of the
  seven known ULXs with luminosities exceeding the Eddington
  luminosity for a $10M\sun$ black hole, a low luminosity active
  galactic nucleus with $2-10\kev$ luminosity of $1.6\times
  10^{39}{\rm~erg~s^{-1}}$, and soft thermal extended emission from
  NGC~5194 detected with \chandra{}. In addition, we also detected a
  new ULX with luminosity $\sim 10^{39}{\rm~erg~s^{-1}}$. We have
  studied the spectral and temporal properties of the LLAGN and 8 ULXs
  in NGC~5194, and an ULX in NGC~5195.  Two ULXs in NGC~5194 show
  evidence for short-term variability, and all but two ULXs vary on
  long time scales (over a baseline of $\sim 2.5{\rm~years}$),
  providing strong evidence that these are accreting sources. One ULX
  in NGC~5194, source 69, shows possible periodic behavior in its
  X-ray flux. We derive a period of $5925\pm200\s$ at a confidence
  level of $95\%$, based on three cycles.  This period is lower than
  the period of $7620\pm500\s$ derived from a \chandra{} observation
  in 2000.  The higher effective area of \xmm{} enables us to identify
  multiple components in the spectra of ULXs. Most ULXs require at
  least two components -- a power law and a soft X-ray excess
  component which is modeled by an optically thin plasma or multicolor
  disk blackbody (MCD).
  However, the soft excess emission, inferred from all ULXs except
  source 69, are unlikely to be physically associated with the ULXs as
  their strengths are comparable to that of the surrounding diffuse
  emission. The soft excess emission of source 69 is well described
  either by a two temperature {\it mekal} plasma or a single
  temperature {\it mekal} plasma ($kT \sim 690\ev$) and an MCD ($kT
  \sim 170\ev$). The MCD component suggests a cooler accretion disks
  compared to that in Galactic X-ray binaries and consistent with that
  expected for intermediate mass black holes (IMBHs).  An iron
  K$\alpha$ line ($EW \sim 700\ev$) or K absorption edge at $\sim
  7.1\kev$ is present in the EPIC PN spectrum of source 26.  The
  spectrum of the ULX in NGC~5195, source 12, is consistent with a
  simple power law.  The LLAGN in NGC~5194 shows an extremely flat
  hard X-ray power-law ($\Gamma \sim 0.7$), a narrow iron K$\alpha$
  line at $6.4\kev$ ($EW\sim 3\kev$), and strong soft X-ray excess
  emission.  The full band spectrum is well described by a two
  component {\it mekal} plasma and reflection from cold material such
  as putative torus.
\end{abstract}


\section{Introduction}
X-ray emission from spiral galaxies consists of several components
including an active nucleus (if present), extra-nuclear point X-ray
sources, hot gas, supernova etc.  Many spiral galaxies contain a
mildly or non-active nucleus and, in this case, the luminosity in the
X-ray band is dominated by the emission from a few luminous
off-nuclear point sources (Roberts \& Warwick 2000). The brightest of
these X-ray sources, with luminosities exceeding the Eddington
luminosity for a $10M\sun$ black hole, are defined as the
ultra-luminous X-ray sources (ULXs).  These objects, which may
sometimes individually outshine the rest of the galaxy, were first
discovered in {\it Einstein} observations (Fabbiano \& Trincheri 1987;
Fabbiano 1989). Later they were detected in large numbers by \rosat{}
(Colbert \& Mushotzky 1999; Roberts \& Warwick 2000; Colbert \& Ptak
2002).  The first detailed X-ray spectral characteristics of ULXs were
revealed by \asca{} and demonstrated that most ULXs exhibit multicolor
disk blackbody spectra with a `high temperature' similar to that
observed from Galactic X-ray binaries in their high state. Some ULXs
were also found to vary and make a transition, at lower luminosities,
to power-law spectra, similar to the spectral transitions seen in many
Galactic XRBs. However, the observed X-ray flux of many ULXs implies
(assuming isotropic emission and Eddington-luminosity limited systems)
black holes of $\sim 10^2 - 10^4 M\sun$. These masses are much larger
than the measured masses of known Galactic X-ray binaries or black
holes (GXRBs or GBHs) and much lower than the measured masses of
active galactic nuclei, thus making them candidates for
intermediate-mass black holes (IMBHs).

Several models have been proposed to explain the high luminosities of
ULXs. These include ($i$) accreting IMBHs with sub-Eddington rates
(e.g., Colbert \& Mushotzky 1999), ($ii$) XRBs with anisotropic
emission (King et al. 2001), ($iii$) beamed XRBs with relativistic
jets directly pointing towards us i. e., scaled down versions of
blazars (Mirabel \& Rodriguez 1999), and ($iv$) XRBs with
super-Eddington accretion rates (Begelman 2002).  A few ULXs have been
found to be young supernova (e.g., SN~1979, Immler, Pietsch \&
Aschenbach 1998) with X-ray emission powered by the interactions of
shock waves and the surrounding dense inter stellar medium. Some of
the accretion-powered ULXs are also observed to be associated with
older supernova remnant-like structures (e.g., IC~342-X1, Roberts et
al. 2003).  The recent observation of a break at a frequency of $\sim
28{\rm~mHz}$ in the power density spectrum of the ULX NGC~4559 X-7
suggests a mass of a few thousands solar masses, consistent with the
measured thermal temperature of $kT = 120 \ev$.  This observation
clearly supports the IMBH scenario for ULXs. It has become
increasingly clear that there are two separate populations of ULXs. A
large number of ULXs are located in regions of current star formation
in star-burst and spiral galaxies (The Antennae, Zezas et al. 2002;
NGC~3256, Lira et al. 2002; NGC~4485/90, Roberts et al. 2002; Arp 299,
Zezas, Ward, \& Murray 2003; M~51, Terashima \& Wilson 2004). These
ULXs are likely to be associated with young stellar populations. Other
ULXs have been observed in early-type elliptical/SO galaxies and are
likely to be associated with older stellar populations. A recent study
of the ULX population of 82 galaxies by Swartz et al. (2004) suggests
that $\sim 25\%$ of these sources may be background objects, including
$14\%$ in spiral galaxies and $44\%$ in elliptical galaxies.

\chandra{} has revolutionized the detection of ULXs primarily due to
its superior angular resolution (see e. g., Swartz et al. 2004).
\xmm{}, on the other hand, has been probing the detailed
characteristics of the ULX populations in nearby galaxies due to its
large effective area. \xmm{} is well suited to study detailed spectral
and temporal characteristics of ULXs in nearby galaxies (e.g.,
Strohmayer \& Mushotzky 2003; Cropper et al. 2003; Miller et al. 2004;
Dewangan et al. 2004).

M~51, also known as the Whirlpool galaxy, is a face-on spiral galaxy
and is located at a distance of 8.4 Mpc (Feldmeier et al. 1997).
Optical studies of emission lines classified it as a LINER or a
Seyfert 2 galaxy (Stauffer 1982).  Furthermore, Ho et al. (1997)
suggest the presence of a broad H$\alpha$ line.  Kohno et al. (1996)
found a nuclear molecular disk, and constrained the dynamical mass
within 70 pc of the nucleus to be $4 - 7\times10^6 M_{\odot}$.  This
indicates that M~51 hosts as massive a black hole as many AGNs.

X-ray observations also strongly suggest the presence of an AGN in
M~51.  {\it Einstein} (Palumbo et al. 1985) and ROSAT (Marston et al.
1995; Ehle et al. 1995) observations constrained the soft X-ray
luminosity of the M~51 nucleus to be $L_{\rm X}<5\times10^{39}$erg
s$^{-1}$.  {\it Ginga} scanning observations in 1988 detected bright
hard X-ray emission with $2-20\kev$ luminosity of
$\sim1.2\times10^{41}{\rm~erg~s^{-1}}$, a photon index of
$1.43\pm0.08$ and an intrinsic absorption of
$<7\times10^{21}$cm$^{-2}$ (Makishima et al. 1990).  Such hard nuclear
X-ray emission is often considered to be evidence for a low luminosity
AGN (LLAGN).  \asca{} observed M~51 in the hard ($2-10\kev$) X-ray
band in 1993, and did not detect such a bright hard component.
Instead, a faint hard X-ray continuum with a neutral Fe~K$\alpha$ line
was detected, whose flux was an order of magnitude lower than that
measured by Ginga (Terashima et al. 1998).  The reason for this large
difference in luminosity has recently been clarified by \sax{}
observations. Fukazawa et al. (2001) find that the nucleus is
photoelectrically absorbed below 10~keV but is observed directly above
20~keV, implying an absorbing column of ${\rm N_H} \simeq
5.6\times10^{24}{\rm~cm^{-2}}$. The 2-10~keV luminosity measured by
{\it BeppoSAX} was similar to that of {\it ASCA}.  Fukazawa et al.
(2001) attributed the higher {\it Ginga} luminosity to variability of
the absorbing column.

High angular resolution \rosat{} HRI observations also revealed eight
X-ray point sources and diffuse soft X-ray emission in M~51.  {\it
  Chandra} observed M~51 during June 2000 and June 2001 with ACIS-S.
These data have been studied in detail by Terashima \& Wilson (2001,
2004). The X-ray image revealed the nucleus, 113 X-ray sources and
extended emission which resembled the morphology of both the radio and
optical emission line images (Terashima \& Wilson 2001). The X-ray
image of the nucleus is well represented by a model consisting of soft
thermal plasma ($kT \simeq 0.5{\rm~keV}$), a very hard continuum and
an Fe~K$\alpha$ emission line at 6.45~keV with an equivalent width of
greater than 2~keV.  The X-ray spectra of the extra-nuclear clouds are
well fitted by a thermal plasma model with $kT \simeq 0.5{\rm~keV}$.
The spectral shape and morphology strongly suggest that the clouds are
shock-heated by a bipolar outflow from the nucleus.  Out of 113
extra-nuclear sources, 9 sources have luminosities exceeding
$10^{39}{\rm~erg~s^{-1}}$ in the 0.5-8~keV band. The number of
extra-nuclear sources in M~51 is much higher than in most other normal
spiral and elliptical galaxies and is similar to galaxies experiencing
star-burst activity. The X-ray spectra of most of the detected sources
are consistent with a power-law spectral form with a photon index
between 1 and 2, while one source has an extremely hard spectrum and
two sources have particularly soft spectra.  The X-ray spectra of
three of the ULXs are consistent with both a power-law and multi-color
blackbody model. One ULX showed a remarkable spectrum with prominent
emission lines.

In this paper we present an \xmm{} EPIC observation of M~51. Utilizing
the large collecting area of \xmm{}, we have studied spectral and
temporal properties of nine ULXs and the LLAGN in M~51.  The paper is
organized as follows. In Sect.~2, we outline the observation and the
data selection. In Sect.~3, 4, and 5, we present the spatial,
temporal, spectral analysis. Sect.~6 describes properties of the
individual ULXs. We discuss the results in Sect.~7, followed by
conclusions in Sect.~8.

\section{Observations and Data selection \label{obs}}
The galaxy M~51 was observed by \xmm{} observatory
\citep{Jansenetal01} on 2003 January 15 for a duration of $\sim
21\ks$.  The EPIC PN \citep{Struderetal01} and MOS
\citep{Turneretal01} cameras were operated in full-frame mode using
the thin filter. The optical monitor (OM) instrument and the
reflection grating spectrometers operated simultaneously with the EPIC
cameras.

The raw PN and MOS events were processed and filtered using the most
recent version of the calibration database and analysis software ({\tt
  SAS v6.0}) available in March 2004. Examination of the background
light curves extracted from source-free regions showed that our
observations was not strongly affected by particle induced flares.
Events in known bad pixels were discarded. For the spectral analysis,
we used events with patterns 0--4 (single and double) and flag zero
for the PN, and patterns 0--12 (similar to \asca{} event grades 0--4)
and flag zero for the MOS.  However, for the temporal analysis, we
used all events with pattern 0--12 for both the PN and MOS cameras.

\section{Spatial analysis}
\begin{figure}
  \centering \includegraphics[height=7cm]{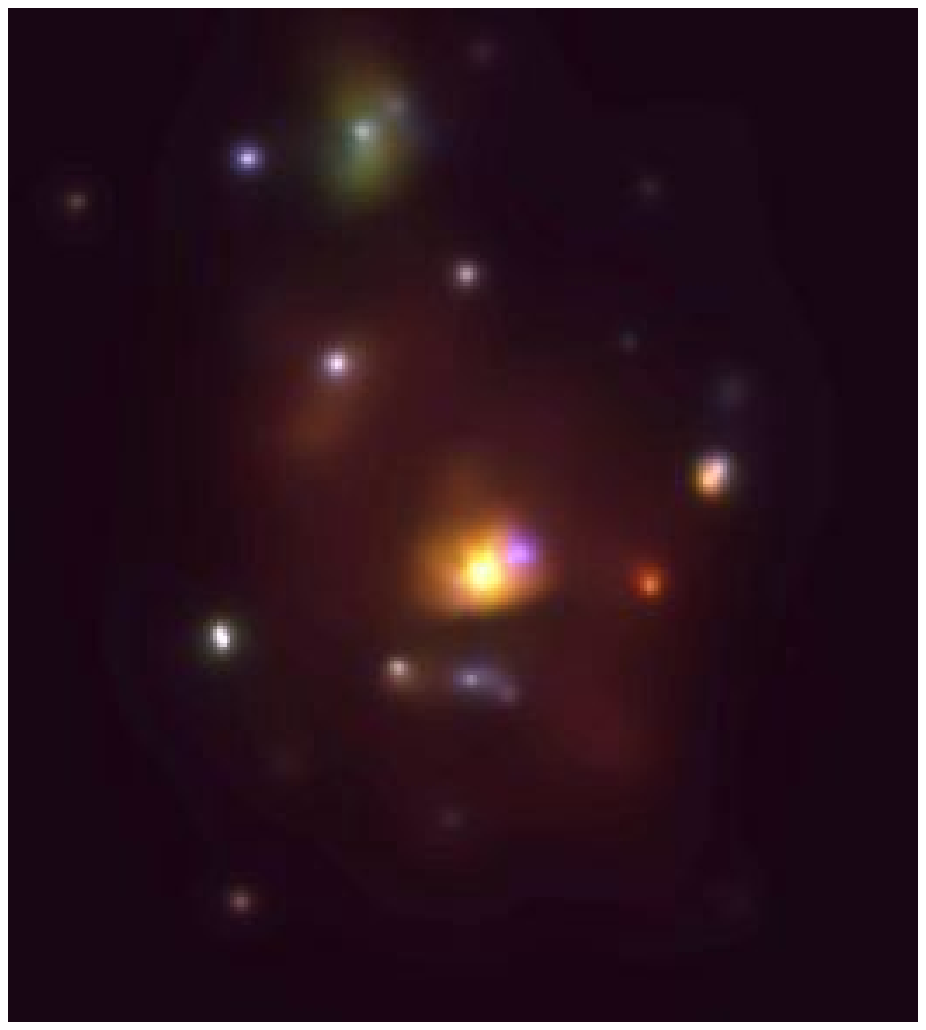}
  \includegraphics[height=7cm]{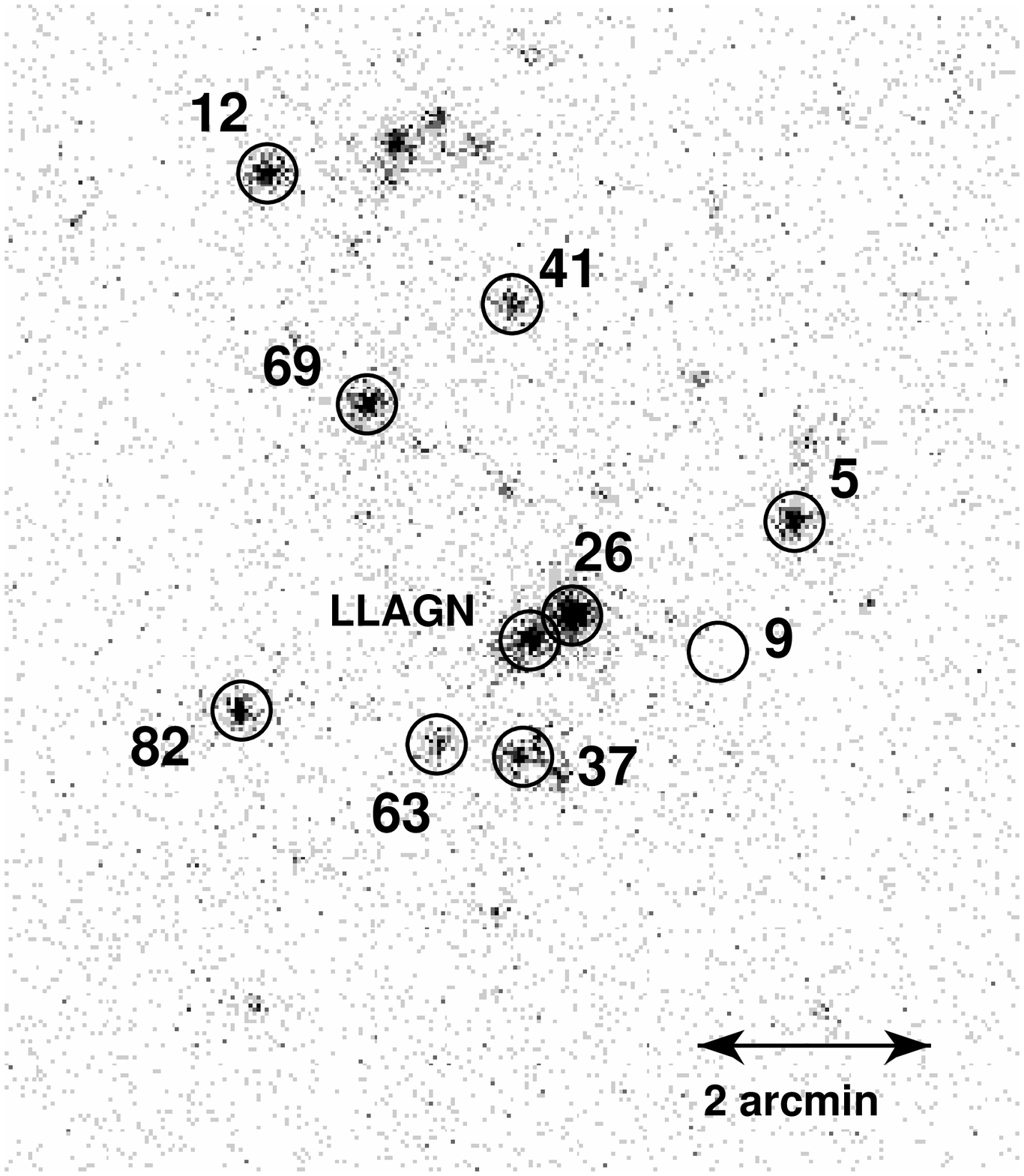}
  \caption{{\it Left:} Three color image of M~51 created from
    adoptively smooth \xmm{} EPIC PN images. The three colors red,
    green, and blue correspond to the $0.2-0.7\kev$, $0.7-2\kev$, and
    $2-10\kev$ energy bands. North is towards the top and east is to
    the left of the image. The extended source at the top of the image
    is the companion galaxy NGC~5195.  The bright extended source near
    the center of the image is the center of the galaxy NGC~5194 that
    consists of extended soft emission and two hard X-ray sources: the
    LLAGN and an ULX NGC~5194 source 26. {\it Right:} The EPIC PN
    image of M~51 in the $2-10\kev$ band showing the positions of 9
    ULXs and the LLAGN.}
  \label{f1}
\end{figure}

\begin{figure}
  \centering \includegraphics[width=10cm]{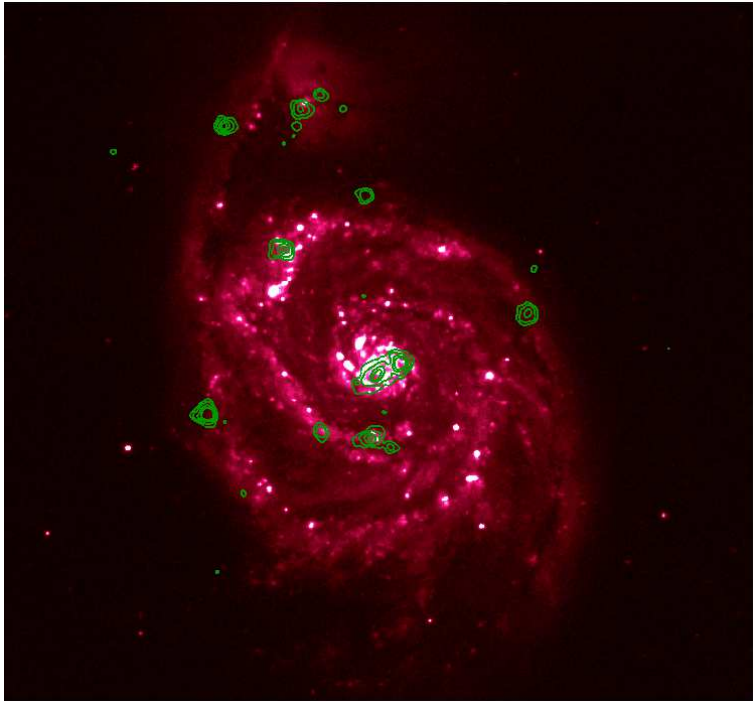}
\caption{\xmm{} ultra-violet image overlaid with the 
  $1.5-10\kev$ band X-ray contours, derived from the PN image, showing the discrete hard X-ray
  sources.  The contours are drawn at levels of $0.4$, $0.7$, $1.3$, $1.9$,  and $2.6{\rm~counts~pixel^{-2}}$ ($1{\rm~pixel} = 0.953{\rm~arcsec}$). The
    image size $11.1\arcmin \times 11.1\arcmin$. North is up and east is
    left. Two central hard X-ray sources, about $20\arcsec$ apart, are clearly resolved by the EPIC
  instruments. The center of UV image coincides with the fainter of the two
  hard X-ray sources in the central regions. This fainter source is the LLAGN
  in NGC~5194. The brighter source is the ULX NGC~5194 source 26.
  \label{f2}}
\end{figure}

We extracted PN images in the full ($0.2-10\kev$), $0.2-0.7\kev$,
$0.7-2\kev$, and $2-10\kev$ bands. First we adaptively smoothed the
full band image using the CIAO task {\tt csmooth} and created a
smoothing kernel. The images in the $0.2-0.7\kev$, $0.7-2\kev$ and
$2-10\kev$ bands were then adaptively smoothed using the kernel
created above, and combined to form a three-color image shown in
Figure~\ref{f1} ({\it Left}).
The color image shows several discrete sources, diffuse soft emission,
and the bright nucleus of NGC~5194.  Interestingly, the bright nucleus
of NGC~5194 actually consists of two hard X-ray sources $\sim
20\arcsec$ apart (see below).  Also present is the companion galaxy
NGC~5195 showing diffuse emission, multiple sources in the central
region, and a bright point source to the left of the nucleus.  To
avoid any possible confusion on multiple source identification numbers
arising from multiple X-ray missions, we follow the \chandra{} source
identification numbers as set by Terashima \& Wilson (2004).
Figure~\ref{f1} ({\it right}) shows the EPIC PN image in the
$2-10\kev$ band and the positions of the ULXs studied in this paper
have been marked.  The ULX source 9 in NGC~5194 is not seen in the
hard band image image due to its location in the chip gap and its
extreme softness.  The source counts and positions listed in Table 1
have been determined by a source detection procedure by utilizing a
series of SAS tasks as described by Miyaji et al. (2003).  In short,
an input image is searched for point-like sources by a simple
sliding-cell detection method with local background using {\em
  eboxdetect}. This is followed by a ``background map" creation by a
spline fit to the source-excluded region using {\em esplinemap}. A
sliding cell detection is repeated to find excess over this background
map ({\em eboxdetect}). The final source positions and counts are
determined by a multi-source maximum-likelihood fit with the \xmm{}
PSF, using the sources detected in the previous step as the starting
points ({\em emldetect}).  In order to be free from the effects of
soft diffuse emission, which could prevent the accurate determination
of the background map, the hard-band (2-10 keV) image is used to
obtain the source counts, except source 9, which has not been detected
in the hard band.  In Figure~\ref{f2}, we show the UV image of M~51
taken with the \xmm{} OM. The image shows beautiful spiral arms, a
prominent nucleus, and bright star clusters surrounding the nucleus
with either a ring-like morphology or multiple inner spiral arms.  One
of the spiral arm of NGC~5194 appears to be tidally connected to the
companion galaxy NGC~5195. This galaxy shows a bright nucleus with
dark dust lanes near the nucleus.
 
In Fig.~\ref{f2}, we have overlaid the X-ray contours onto the UV
image.  The contours represent the hard ($1.5-12\kev$) X-ray sources
only.  Fig.~\ref{f2} highlights the two bright hard X-ray sources in
the central region of NGC~5194.  The two sources are separated by
$\sim 20\arcsec$.  The fainter of the two X-ray sources coincides with
the UV nucleus. The brighter source is an ULX, which is clearly
resolved in \chandra{} images (Terashima \& Wilson 2004).

In this paper we study the active nucleus and the point sources with
luminosities $\ge 10^{39}{\rm~erg~s^{-1}}$ in the $0.3-12\kev$ band.
There are eight ULXs in addition to the LLAGN in NGC~5194. There is
also a well isolated ULX in the companion galaxy NGC~5195. However,
the central region of NGC~5195 consists of multiple bright point
sources and extended soft emission as seen in the \chandra{} ACIS-S
image (Terashima \& Wilson 2004).  Due to lower angular resolution of
\xmm{}, it is not possible to extract counts for individual sources in
the central region of NGC~5195. Therefore we exclude the bright
sources in the central region of NGC~5195 from our analysis.  The
\chandra{} source NGC~5194\#63 had an X-ray luminosity $<
10^{39}{\rm~erg~s^{-1}}$ during the two previous \chandra{}
observations. In this observation this source has a luminosity of
$\sim 2\times 10^{39}{\rm~erg~s^{-1}}$ in the $0.3-10\kev$ band and we
now include this source in our sample.

\section{Temporal analysis}


\subsection{Short-term variability}
We extracted $0.2-12\kev$ X-ray light curves for the LLAGN and ULXs in
M~51 from the EPIC PN data. The extraction regions had radii in the
range of $15-35\arcsec$ and enclosed $>60\%$ of the X-ray emission
from each source.  The smallest circular regions were chosen to avoid
contamination from nearby sources.  We used time bins of $520\s$ and
required that the bins be at least $50\%$ exposed in order to minimize
the errors due to the low number of counts from the weak sources. We
also extracted individual background light curves for each of the
sources using circular regions lying approximately at the same readout
distance as that for the source. The source light curves were then
corrected for the background contribution using the background light
curves and appropriate scaling factors to take into account the
different extraction areas.

\begin{figure}
  \centering \includegraphics[width=7cm,angle=-90]{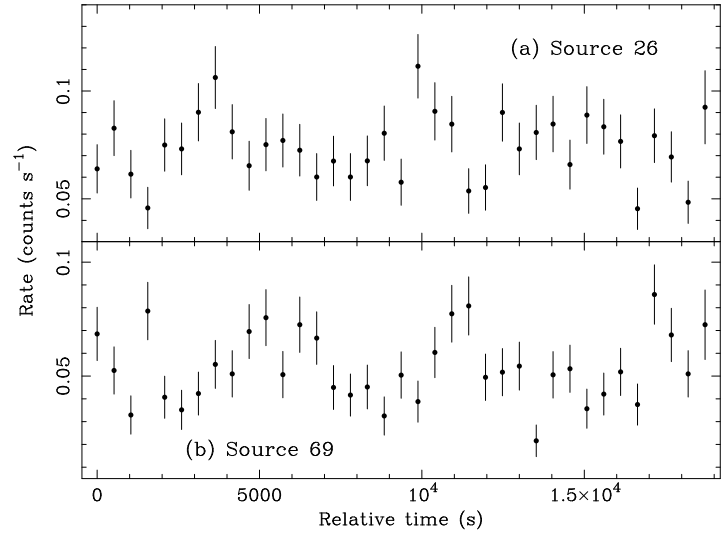}
  \caption{Short-term EPIC PN light curves of ULXs in M~51 (a) source
    26, and (b) source 69. The bin sizes are $520\s$. The light curves
    have been corrected for the background contribution. }
  \label{f3}
\end{figure}

We fit a constant count rate model to each of the light curves to
investigate the possibility of rapid variability. We rebin the light
curves, if required, to ensure a minimum of 20 counts per bin to allow
the use of $\chi^2$ fitting statistics. The results for each light
curve are listed in Table~2. We find that only two of the sources show
significant variability: source 26 (at the $98.6\%$ confidence level)
and source 69 ($99.95\%$ confidence level). The light curves of these
two sources are plotted in Figure~\ref{f3}.


\subsection{Periodicity}
\begin{figure}
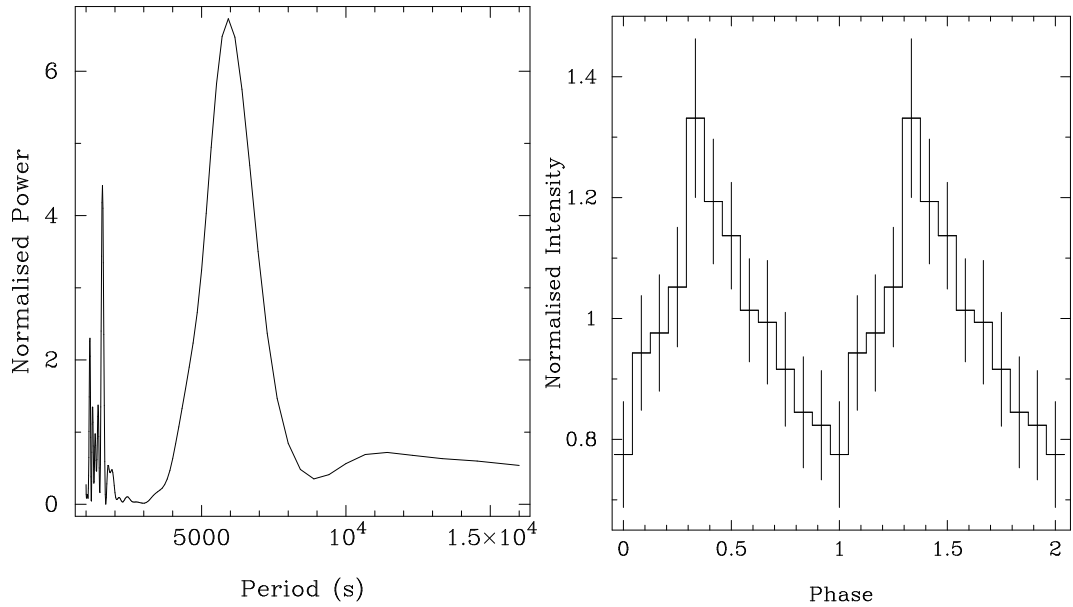

  \centering \includegraphics[width=7cm]{f4a.ps}
  \includegraphics[width=7cm]{f4b.ps}
  \caption{{\it Left:} Periodogram of NGC~5194 source 69. {\it Right:}
    Folded \xmm{} light curve of NGC~5194 source 69.}
  \label{f4}
\end{figure}

Source 69 shows the most significant rapid variability. The light
curve shows three tentative peaks at regular intervals. The first
\chandra{} observation of this ULX in June 2000 showed two peaks
separated by $\sim 7000\s$. These features were interpreted by Liu et
al. (2002) as a signal with a period of $\sim 2.1{\rm~hr}$. We have
searched for a similar periodic signal in the \xmm{} light curves. For
this purpose, we combined the PN and MOS light curves, binned to
$520\s$. The initial search for periodicity in the source, as well as
the folding of the light curve with respect to the final period
obtained was performed with the XRONOS 5.19 package, which is a part
of FTOOLS 5.2.  The periodogram was obtained using the algorithm of
Scargle (1982), following the recipe of normalizing the periodogram
power using the total variance of the complete light curve as
prescribed by Horne \& Baliunas (1986). This algorithm is very robust
for detecting periodic signals in light curves with few data points as
well as non-uniformly distributed observational data, while providing
a `false alarm probability' of the periodicity arising in the light
curve from random noise. This probability is obtained using the signal
to noise ratio, where the variance of the noise is obtained under the
basic assumption that it is a normally distributed random variable.
The periodogram is shown in Figure \ref{f4} ({\it left}). The period
obtained was $5925\pm200$ s, with a 'false alarm probability' of
$0.045$, suggesting that the observed modulation is not generated by
the statistical fluctuation of the noise, to a confidence level of
$95.5\%$. The main source of the noise in the power in the Fourier
frequency domain are the inherent statistical fluctuations at scales
$< 2$ ks (Figure \ref{f4}).  We have tested for the periodicity on top
of a constant count rate. However, in reality, the source probably has
a red-noise power spectrum, similar to XRBs. Therefore, the $\sim
2\sigma$ or $95.5\%$ detection of the period is an upper limit, and
the periodicity may or may not be a true signal.
Future long baseline X-ray observations will be crucial for the robust
detection of the periodicity from this ULX.  In Figure \ref{f4} ({\it
  right}) we provide the light curve folded with respect to the period
of 5925 s, with a bin size of $520\s$. Evidently, the modulation
fraction is in excess of $45\%$, similar to the estimate of Liu et al.
(2002).

\subsection{Long-term Variability}
\begin{figure}
  \centering \includegraphics[width=15cm]{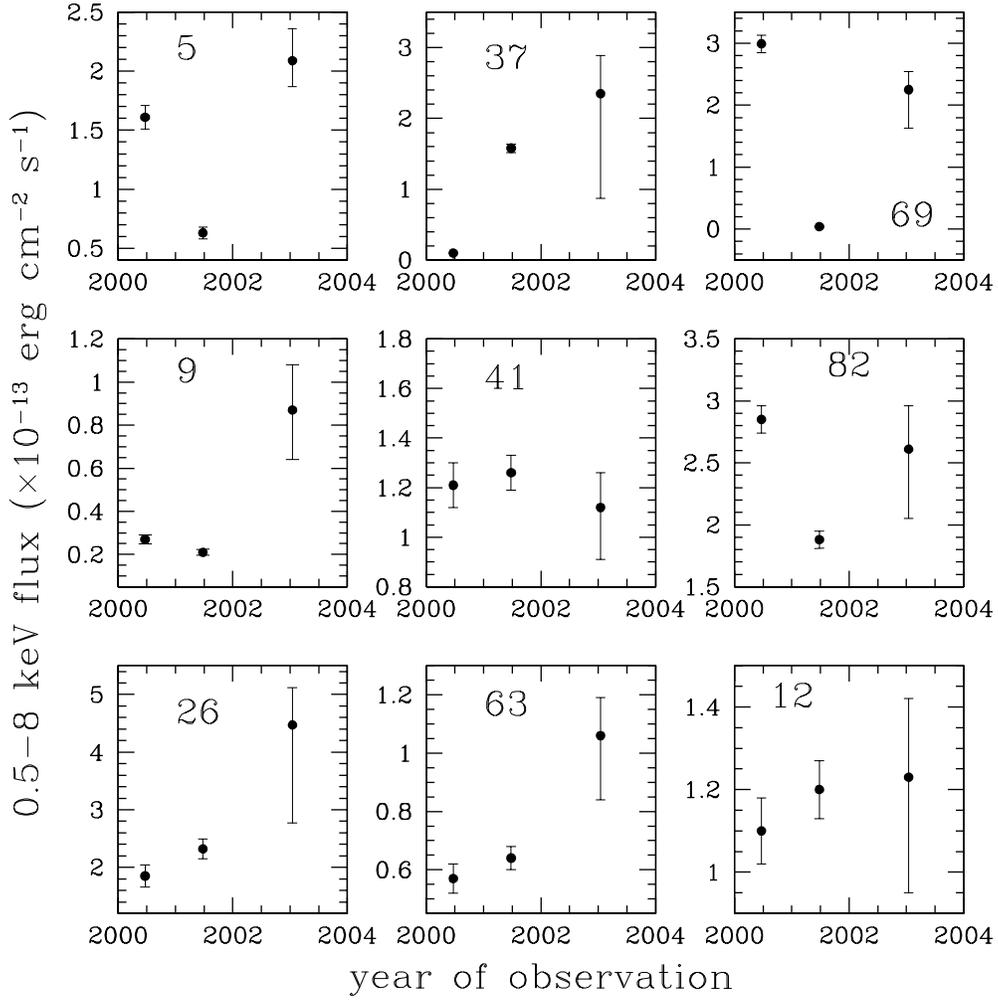}
  \caption{Long-term variability of ULXs in M~51 over a baseline of
    two and half years. The fluxes for the years 2000 and 2001 were
    derived from the two \chandra{} observations by Terashima \&
    Wilson (2004). All but two ULXs 41 and 12 clearly varied.}
  \label{f5}
\end{figure}

We have compared the $0.5-8\kev$ fluxes derived from the \xmm{}
observation (see below) and earlier two observations with \chandra{}
(Terashima \& Wilson 2004) in June 2000 and June 2001. Figure~\ref{f5}
shows the comparison of the fluxes. Cross-calibration results suggest
that the fluxes measured with \chandra{} and \xmm{} agree at a level
better than $10\%$ (Snowden 2002). Thus, all the ULXs, except
NGC~5194\#41 and NGC~5195\#12, show long-term significant variability
over a baseline of $2.5{\rm~yr}$.

\section{Spectral Analysis \label{analysis}}
For each ULX and the LLAGN, we extracted source and background spectra
from the final filtered PN event list using the same extraction
regions that we used for the PN light curves. We also extracted the
source spectra from the MOS data using source regions similar to that
used for the PN. The background MOS spectra for the individual ULXs
were extracted using circles in the nearby source-free regions.
Appropriate PN and MOS responses and effective area files were created
for each source using the SAS tasks {\it rmfgen} and {\it arfgen}. A
grouping of 20 counts per spectral channel resulted in significant
degradation in the spectral resolution of the ULX spectra above
$2kev$. Therefore, all the source spectra were grouped to a minimum of
10 counts per spectral channel so that discrete spectral features are
not missed.  The spectra were analyzed with the {\tt XSPEC 11.3}
spectral analysis package (Arnaud 1996), and using the C-statistics
instead of $\chi^2$-statistic which is not useful in the case of low
($\ltsim 20$) counts per bin. Unlike the $\chi^2$-statistic, the
C-statistic does not provide a goodness-of-fit (GOF) criterion.
Therefore we simulate 5000 spectra based on the best-fit model derived
using the C-statistic, and calculate the percentage of these spectra
with the C-statistic less than that for the data.  This number
provides the GOF, and if this number is $\sim 50\%$, then the model is
a good fit to the observed spectrum (similar to the reduced $\chi^2
\sim 1$ criterion), though obviously we cannot statistically exclude
fits for which GOF$\le 95\%$.  The confidence intervals are calculated
in the same way as in the case of $\chi^2$-statistic. All the errors
quoted below were calculated at $90\%$ confidence level for one
interesting parameter i.e., $\Delta C = 2.7$.

At first the MOS and PN spectra of a few bright sources were fitted
separately to check for the possible uncertainties due to
cross-calibration problems. We found generally good agreement between
MOS and PN cameras in the $0.3-10\kev$ bands.  Therefore we present
the spectral results obtained by fitting the same model jointly to the
PN and MOS data while leaving the relative normalizations to vary.

\subsection{The low luminosity AGN in NGC~5194}
\begin{figure}
  \centering \includegraphics[width=5cm,angle=-90]{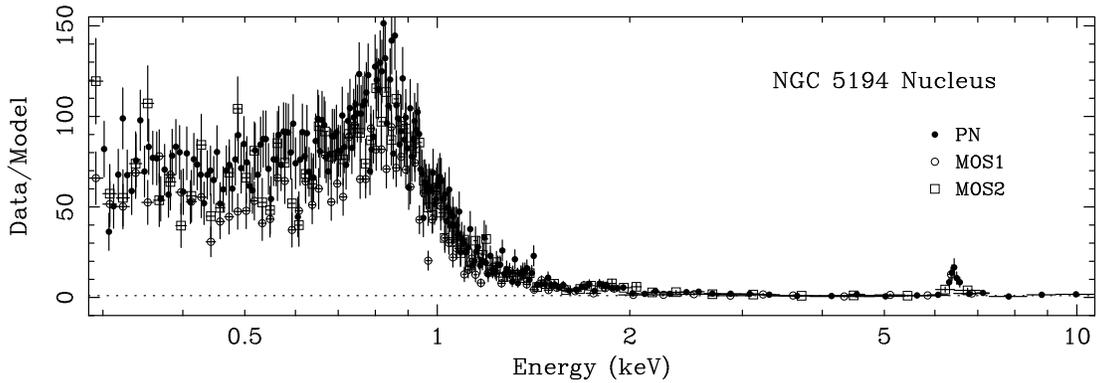}
\caption{Ratio of the observed EPIC data and the best-fit power-law
  model obtained by fitting the data in the $3-6\kev$ and $7-10\kev$ bands.
  \label{f6}}
\end{figure}

\begin{figure}
  \centering \includegraphics[width=8cm,angle=-90]{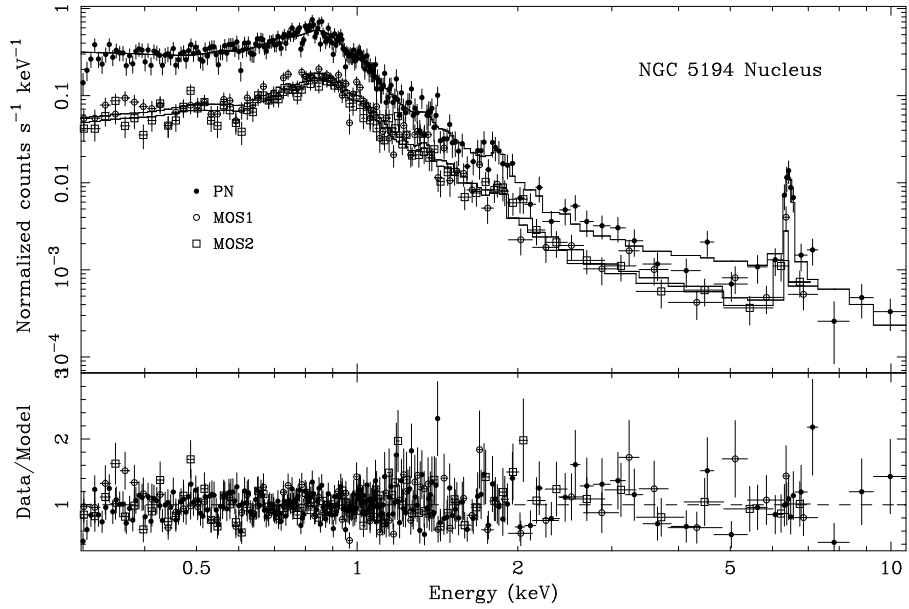}
\caption{The observed EPIC PN and MOS spectra of the nucleus of NGC 5194 and
  the best-fitting spectral model which a combination of two {\it mekal} plasma
  components, power-law, and a
  narrow  Gaussian line at $6.4\kev$ and multiplied by the Galactic absorption.
  \label{f7}}
\end{figure}

An absorbed power-law model is a poor fit to the PN and MOS spectra of
the nucleus of NGC~5194 ($C= 4117.2$ for 409 degrees of freedom (dof),
GOF=$100.0\%$). To show the significant spectral features, we fitted
the absorbed power-law model in the $3-5\kev$ and $7-10\kev$ bands and
plotted the ratio of the observed data in the $0.3-10\kev$ band and
the best-fit power-law in Figure~\ref{f6}.  A strong soft X-ray excess
below $1.5 \kev$ and a narrow iron K$\alpha$ line at $6.4\kev$ are
evident in the spectrum.  The X-ray image, shown in Fig.~\ref{f1},
clearly shows presence of extended emission which we interpret as
originating from optically thin thermal plasma.  Accordingly, we added
a {\it mekal} component to describe the soft component and a Gaussian
line model at $6.4\kev$. The fit improves significantly ($C= 452.3$
for 404 dof, $GOF= 88.5\%$), however, the soft X-ray emission is not
well described. There are still line-like residuals below $2\kev$,
suggesting either the hot gas responsible for the soft excess emission
has a temperature structure and/or its abundance ratio is non-solar.
Including an additional {\it mekal} component (model 1) further
improves the fit ($C=414.7$ for 401 dof; $47.8\%$), and removes the
soft residuals seen earlier. The best-fit temperatures are
$kT=171_{-29}^{+23}\ev$ and $kT=610_{-14}^{+16}\ev$, and the best-fit
abundances, relative to solar, are $0.20_{-0.14}^{+2.49}$ and
$0.18_{-0.02}^{+0.05}$ for the two {\it mekal} components.  The
observed data, the best-fit model and their ratio are plotted in
Figure~\ref{f7}.  We also tested the possibility of the non-solar
abundance ratio by replacing the two {\it mekal} component by a {\it
  vmekal} component (model 2) which allows the abundance ratios of
individual elements to vary. This model also improved the fit over
that involving a single {\it mekal} component. The best-fit parameters
for both models are listed in Table~\ref{t3}.  Model 1 and 2 describe
the data statistically equally well ($GOF=48\%$, and $47\%$ for model
1 and 2, respectively).  Both models result in an extremely flat hard
X-ray slope ($< 0.8$), and strong iron K$\alpha$ line (EW $\sim
3\kev$) from neutral material.

The flat spectrum and strong iron line strongly suggest that the hard
X-ray emission from the LLAGN in NGC~5194 is dominated by reflection
from a cold material. Previous \sax{} observations of M~51 showed the
LLAGN is obviously Thomson thick ($N_H = 5.6_{-1.6}^{+4.0}\times
10^{24}{\rm~cm^{-2}}$; Fukazawa et al. 2001). Therefore, we replaced
the simple power-law model with the Compton reflection model {\it
  pexrav} (Magdziarz \& Zdziarski 1995). This model calculates the
reflected spectrum from a neutral disk exposed to an exponentially
cutoff power-law spectrum. Since \xmm{} does not cover the expected
peak (20--40 keV) of the reflection component, it is not possible to
tightly constrain all the parameters of the reflection model using
this data alone. Instead, we fix the power-law photon index at $1.9$,
cutoff energy of the primary power law at 200 keV, disk inclination at
70 degrees and the abundance of heavy elements at $0.1$ relative to
solar value. We also fixed the relative amount of reflection, $R = 1$,
corresponding to an isotropic source above a reflecting plane. The
only variable parameter of the pexriv model was the normalization of
the incident power law. We chose a high inclination angle due to the
Thomson-thick nature of the LLAGN. We also set the reflection model to
produce the reflection component alone.  This corresponds to a
situation in which the primary X-ray emission is completely absorbed
by a column density in excess of $\sim 10^{24}{\rm~cm^{-2}}$ along the
line of sight and we observe the X-ray emission scattered by the
matter near the nucleus.  We also include a narrow Gaussian line model
to describe the Fe K emission.  Fitting the above model (model 3)
resulted in no strong apparent residuals.  Model 3 consisting of two
{\it mekal} components, Compton reflection, and a narrow Gaussian
resulted in $C = 420.9$ for 400 dof and $GOF = 61.8\%$. We consider
this model to be the best-fit model as it is more physical than th
earlier model involving simple power-law. The best-fit {\it mekal}
temperatures are $177_{-32}^{+36}\ev$, $614_{-11}^{+13}\ev$.  The iron
line equivalent width is $2.5_{-1.4}^{+1.4}\kev$, similar to that
measured from \chandra{} observations (Levenson et al. 2002).  The
best-fit parameters are listed in Table~\ref{t3} (Model 3). The
observed flux is $8.2\times 10^{-13}{\rm~erg~s^{-1}}$ in the
$0.3-10\kev$ band, and $2.0\times 10^{-13}{\rm~erg~s^{-1}}$ in the
$2-10\kev$ band, which correspond to luminosities $6.9\times
10^{39}{\rm~erg~s^{-1}}$ in the $0.3-10\kev$ band, and $1.7\times
10^{39}{\rm~erg~s^{-1}}$, respectively.

\subsection{ULXs in M~51}
\begin{figure}
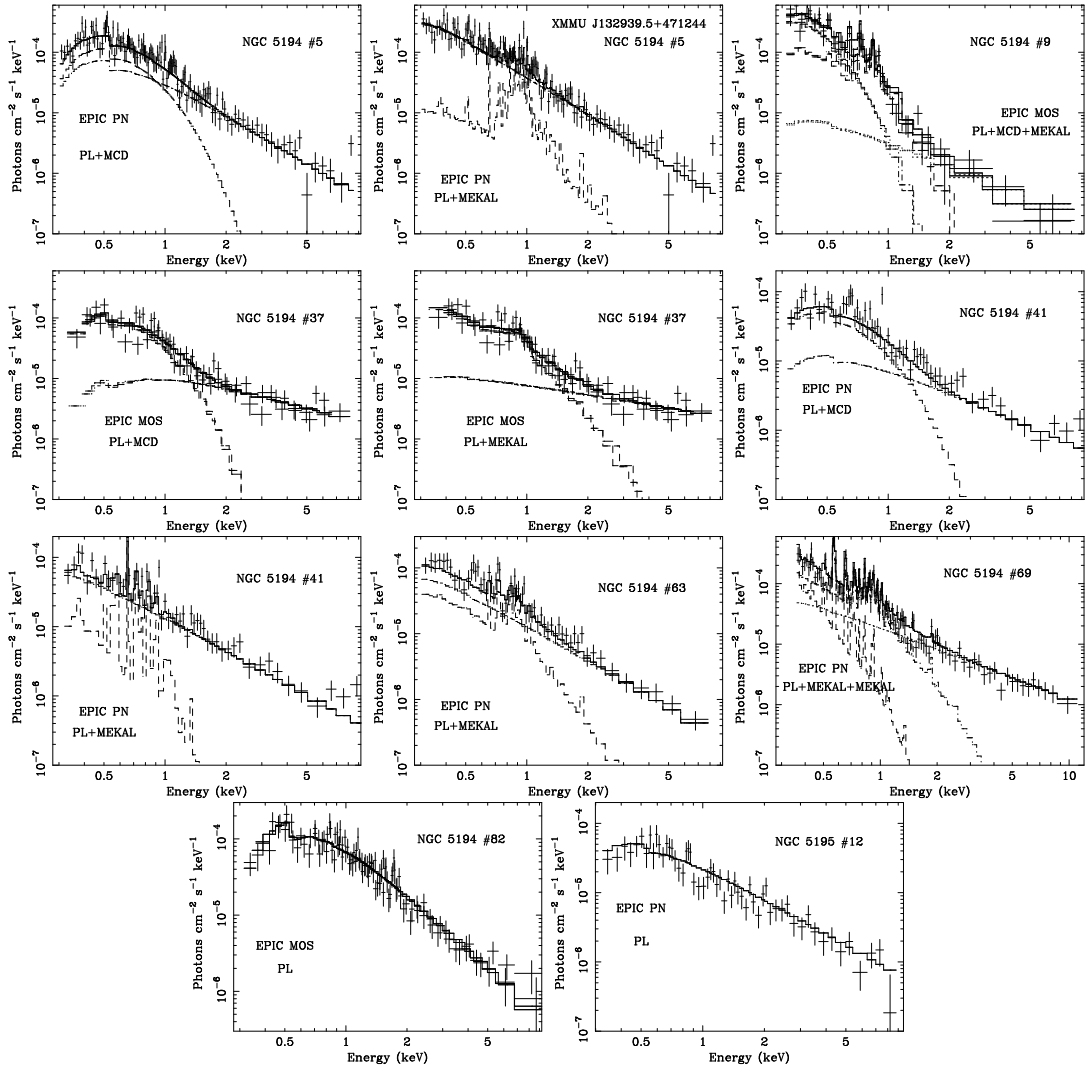

  \centering \includegraphics[width=3.5cm,angle=-90]{f8a.ps}
  \includegraphics[width=3.5cm,angle=-90]{f8b.ps}
  \includegraphics[width=3.5cm,angle=-90]{f8c.ps}
  \includegraphics[width=3.5cm,angle=-90]{f8d.ps}
  \includegraphics[width=3.5cm,angle=-90]{f8e.ps}
  \includegraphics[width=3.5cm,angle=-90]{f8f.ps}
  \includegraphics[width=3.5cm,angle=-90]{f8g.ps}
  \includegraphics[width=3.5cm,angle=-90]{f8h.ps}
  \includegraphics[width=3.5cm,angle=-90]{f8i.ps}
  \includegraphics[width=3.5cm,angle=-90]{f8j.ps}
  \includegraphics[width=3.5cm,angle=-90]{f8k.ps}
\caption{Unfolded EPIC PN or MOS spectra of ULXs in M51.
  \label{f8}}
\end{figure}

We performed joint fits for the PN and MOS spectra of each ULX with
the following spectral models: ($i$) a power law (PL), ($ii$)
multicolor accretion disk blackbody (MCD), ($iii$) thermal plasma
({\it mekal}), ($iv$) PL+MCD, ($v$) PL+{\it mekal}.  The models also
incorporate photoelectric absorption ({\it wabs} with a minimum value
of $1.5\times 10^{20}{\rm~cm^{-2}}$, the Galactic absorption in the
direction of M~51; Stark et al. 1992). We assume an abundance of $0.1$
relative to the solar in all the {\it mekal} components. The point
spread function for the two sources 82 and 37 partially coincide with
PN CCD chip gaps and/or bad pixel columns, prompting us to restrict
ourselves to using only MOS data for the spectral analysis of these
sources. One or more of the above models provided a reasonably good
fit to all the ULXs except for sources 9, 26 and 69.  The best-fit
models and the unfolded X-ray spectra of these ULXs are shown in
Figure~\ref{f8}.  The best-fit parameters are listed in Table~4.  We
have also listed the $0.5-8\kev$ observed flux and luminosity for the
PN data in Table~4. The errors on flux were estimated at $90\%$ level
based on 1000 sets of parameter values drawn from the distribution
that is assumed to be multivariate Gaussian centered on the best-fit
parameters with sigmas from the covariance matrix. The fluxes measured
with PN and MOS cameras are within the $90\%$ errors quoted in
Table~4. The {\it mekal} model alone does not provide a good fit to
the spectra of all ULXs except source 12, therefore the best-fit
parameters are not listed in Table~4.

X-ray spectra of sources 5, 37 and 41 in NGC~5194 are equally well
described by a PL+MCD or a PL+{\it mekal} model, while source 63
prefers PL+{\it mekal} model. Two ULXs source 82 in NGC~5194 and
source 12 in NGC~5195 only requires a single spectral component.
Source 82 is described by a simple power law and does not
statistically require any soft component at a level $\ge 99\%$. Source
12 is well described either by a simple power law or a {\it mekal}
model. In Table~4, we have listed the reduction in the C-statistic
value ($\Delta C$) for $\Delta p$ additional parameters for the
addition of a soft component ({\it mekal}, MCD or both, see below) to
the simple power-law model. We have also listed the statistical
significance derived from the maximum likelihood ratio (MLR) test. The
$99\%$ criterian for selecting the more complex model is significance
$< 0.01$. As a caveat we note that the MLR test is statistically not
robust in cases that involve testing a hypothesis that is on the
boundary of the parameter space e.g., the null values of the
normalizations of MCD, {\it mekal} and Gaussian lines occur in the
simple power law model (Protassov et al. 2002). The results of the MLR
test presented here should be treated only as indicative.

\par The spectra of sources 9, 26 and 69 are not satisfactorily
described by any of the four models mentioned above. Source 9 is the
softest ULX in our sample. Although not a good fit, the PL+MCD model
is marginally better than the PL+{\it mekal} model for this ULX (see
Table 4). Adding a {\it mekal} component to the PL+MCD model improves
the fit ($\Delta C = -9.7$ for two additional parameters), and results
in a good fit ($C=59.8$ for 50 dof, $GOF=46.9\%$). Replacing the MCD
component by another {\it mekal} component in the PL+MCD+{\it mekal}
model slightly worsened the fit ($C=67.9$ for 51 dof; $GOF=78.2\%$).

\par
\begin{figure}
  \centering \includegraphics[width=4cm,angle=-90]{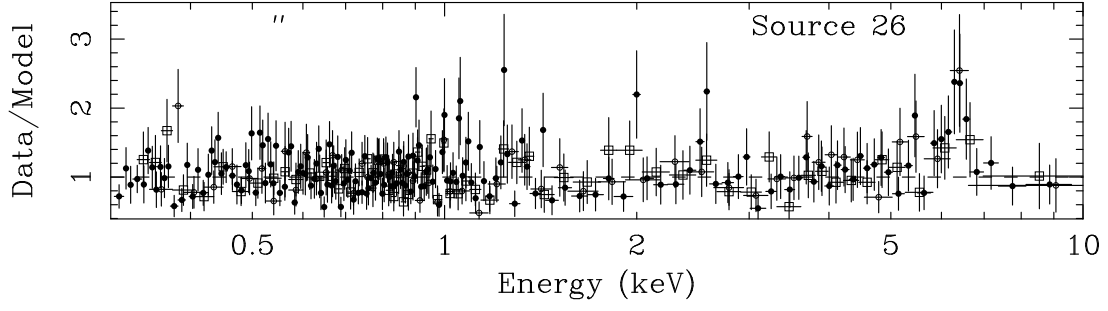}
  \caption{The ratio of the observed EPIC data and the best-fitting
    model consisting of three {\it mekal} components and a power law
    modified by the Galactic absorption. An iron K$\alpha$ line at
    $6.4\kev$ is evident. PN, MOS1, and MOS2 data are marked by filled
    circles, open circles, and open squares, respectively.}
  \label{f9}
\end{figure}

\begin{figure}
  \centering \includegraphics[width=8cm,angle=-90]{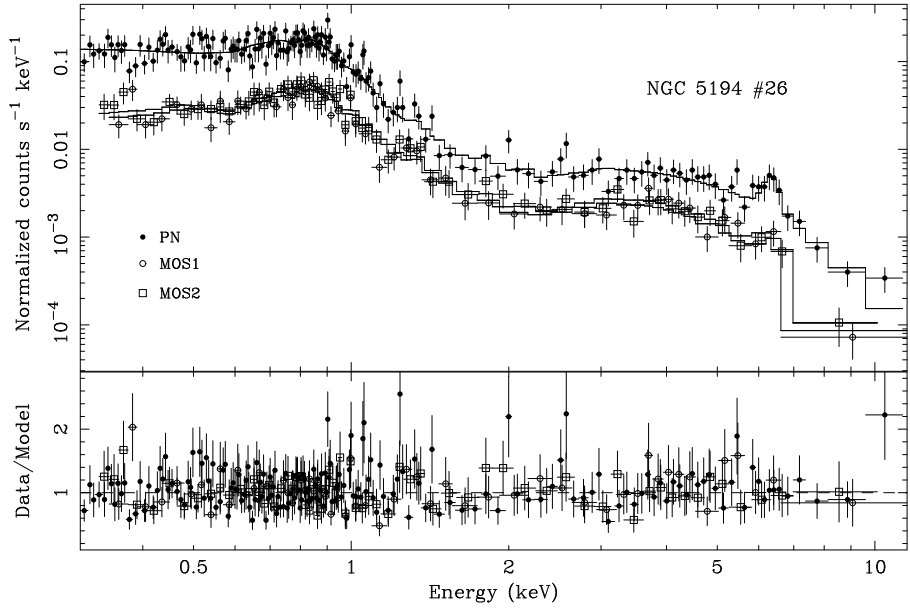}
\caption{The observed EPIC PN and MOS spectra of the ULX NGC~5194\#26 
and the best-fitting spectral model -- a combination of two {\it mekal} plasma
  components,  a Gaussian line at $\sim 6.4\kev$, and an absorbed power-law.
  \label{f10}}
\end{figure}

 The ULX source 26 is located in the region of strong, extended
soft X-ray emission. Its X-ray spectrum is not well described either
by a simple power law or an optical thin plasma.  Similar to the
spectral modeling of the LLAGN, we used multiple {\it mekal}
components to describe the soft emission. Two {\it mekal} components
are statistically required to explain the soft X-ray excess emission.
The addition of the second {\it mekal} component is significant at a
level $>99.99\%$ ($\Delta C = -28$ for two additional parameters). In
Figure~\ref{f9}, we show the ratio of the EPIC data and the best-fit
model consisting of two {\it mekal} components and an absorbed
power-law. An iron K$\alpha$ line at $\sim 6.4\kev$ is clearly seen in
the spectrum. The addition of a Gaussian line improves the fit
($\Delta C = 19.2$ for three additional parameters). The best-fit
resulted in $C=278$ for 295 dof ($GOF = 9.3\%$). The observed EPIC
data, the best-fit model and their ratio are plotted in
Figure~\ref{f10}. The best-fit parameters are listed in
Table~\ref{t5}.

The above fit clearly shows the presence of a Gaussian line. The
best-fit line parameters are $E_{line} = 6.33_{-0.13}^{+0.11}\kev$,
$\sigma = 158_{-155}^{+180}\ev$, line flux $f_{line} =
2.3_{-1.2}^{+1.5}\times 10^{-6}{\rm~photons~cm^{-2}~s^{-1}}$ and $EW =
550\ev$, consistent with an iron K$\alpha$ line from neutral material.
The LLAGN in NGC~5194, a Compton-thick AGN, also shows strong ($EW
\sim 3\kev$; see also Levenson et al. 2002). The ULX NGC~5194\#26 is
separated by only $28\arcsec$ from the LLAGN. The fractional encircled
energy for the PN is $\sim 80\%$ at a redius of $28\arcsec$. The
location of NGC~5194\#26 within the wings of the point spread function
and the very strong narrow iron line of LLAGN raise doubt if the iron
K$\alpha$ emission inferred from the ULX is simply cross-contamination
from the LLAGN. To investigate this possibility, we extracted a
background spectrum using circular regions with centers at the same
distance from the LLAGN as that of NGC~5194\#26. The radii of the two
circles were chosen to be the same as that of the circular region used
to extract the spectrum of NGC~5194\#26. We used the PN data only and
carried out the spectral analysis in the $2-10\kev$ band.  A simple
absorbed power-law model provided $C=83.8$ for 77 dof and $GOF =
57.0\%$. Addition of a Gaussian line improved the fit ($\Delta C = -
13.4$ for three additional parameter). The best-fit parameters are
$\Gamma = 2.4_{-0.5}^{+0.5}$, $n_{PL}=3.0_{-1.8}^{+0.5}\times
10^{-4}{\rm~photons~cm^{-2}~s^{-1}~keV^{-1}}$, $E_{line} =
6.3_{-0.2}^{+0.2}\kev$, $\sigma = 210_{-140}^{+380}\ev$, $f_{line} =
2.7_{-1.5}^{+2.7}\times 10^{-6}{\rm~photons~cm^{-2}~s^{-1}}$,
$EW=690\ev$.  Replacing the Gaussian line with an accretion disk line
(the {\it diskline} model in {\tt XSPEC}) resulted in $C = 69.5$ for
74 dof. The best-fit parameters are $E_{line} =
6.3_{-0.4}^{+0.3}\kev$, disk inclination angle, $i
=26_{-26}^{+32}{\rm~degree}$, $f_{line} = 3.9_{-1.9}^{+2.3}\times
10^{-6}{\rm~photons~cm^{-2}~s^{-1}}$. The emissivity index, inner and
outer radii were kept fixed at $-2.5$, $6r_g$ and $1000r_g$,
respectively, where $r_g$ is the gravitational radius. The addition of
the diskline to the simple power law model improved the fit at a
significance level of $99.7\%$.  We also tested if the broad iron
line-like feature could be described in terms of an absorption edge.
Addition of an edge at $\sim 7.1\kev$ to the absorbed power-law model
improved the fit ($C = 70.6$ for 75 dof). The best-fit parameters are
$N_H= 2.5_{-2.3}^{+2.1}\times 10^{22}{\rm~cm^{-2}}$, $\Gamma =
1.2_{-0.6}^{+0.6}$, edge energy $E_{edge} = 7.1_{-0.1}^{+0.2}\kev$ and
$\tau = 1.4_{-0.6}^{+0.8}$.  Thus the $2-10\kev$ EPIC PN spectrum of
source 26 is equally well described by an absorbed power law and a
broad iron line at $\sim 6.3\kev$ or an iron K-edge at $\sim 7.1\kev$.
It is still possible that the background correction, we adopted here,
may not be accurate due to poor signal-to-noise of the data, the
presence of extended X-ray emission and moderate spatial resolution of
\xmm{}.  Future long X-ray observations with \chandra{} and \xmm{}
will help detecting the line and/or edge unambiguously.

Source 69 is located in a region of extended soft X-ray emission in an
spiral arm.  It is likely that X-ray emission from source 69 is
contaminated by the extended emission. Adding another {\it mekal}
component to the PL+{\it mekal} model improved the fit significantly
($\Delta C = 52.7$ for two additional parameters; $C = 191.9$ for 200
dof; $GOF=18\%$). The best-fit parameters are listed in Table~4.
Replacing one of the {\it mekal} component by the MCD component
slightly worsened the fit ($C=200.9$ for $199$ dof, $GOF=29.2\%$).
However, both the fits are acceptable. The spectral data and the
best-fit model PL+{\it mekal}+{\it mekal} are plotted in
Figure~\ref{f8}.

We compared the soft excess flux in the $0.3-2\kev$ band, modeled as
{\it mekal} component above a power-law, of ULXs with that of the
diffuse emission in the surrounding regions. We estimated the flux of
the diffuse emission from the count rates measured in annular regions
centered at the source positions or nearby circular regions. We used
the best-fit {\it mekal} temperature, estimated from the PL+{\it
  mekal} model for an ULX, to convert the appropriately scaled count
rate of the nearby diffuse emission into the flux in the $0.3-2\kev$.
For all ULXs except source 69, the nearby diffuse emission is
comparable to the soft excess emission, suggesting that the soft
excess emission may not be associated with these ULXs .
The soft excess emission of source 69 ($f_{mekal} \sim
6.6\times10^{-14}{\rm~erg~cm^{-2}~s^{-1}}$) is about a factor of four
stronger than the surrounding diffuse emission in the $0.3-2\kev$
band. Thus, at least a part of the soft excess emission is intrinsic
to the ULX.

\subsubsection{The comptonized accretion disk model}
The power law and the MCD model used to fit the ULX spectra above are
only a mathematical approximation of the real physical spectra and do
not provide much physical insight. Therefore, we also tried a spectral
model based on `real' physics of an accretion disk-corona system. This
model called as the `comptonized accretion disk (CMCD)' has been used
to describe the spectra of six bright ULXs (Wang et al. 2004). The
CMCD model was constructed by Yao et al. (2003). The model assumes a
thermal energy distribution for electrons in a spherical corona around
an accretion disk. The parameters of the CMCD model are the radius of
the spherical corona ($R_c$), electron temperature ($T_c$), coronal
optical depth ($\tau_c$), the temperature of the inner accretion disk
($kT_{in}$), and disk inclination angle ($i$). The model is
implemented as a standard XSPEC table model (see Yao et al. 2003 and
Wang et al. 2004 for more details).

We fitted the CMCD model to the spectra of all the ULXs. It was not
possible to constrain the parameters for all but one ULX source 26,
due to poor signal-to-noise and/or best-fit parameters outside the
tabulated ranges.  Due to the contamination of the extended soft X-ray
emission, source 26 statistically required two {\it mekal} components
in addition to an absorbed CMCD model. The addition of the first and
second {\it mekal} components improved the fit by $\Delta C = -491.1$
and $-19.1$, respectively, for three additional parameters in each
case. The {\it mekal} temperatures are similar to that derived earlier
using the combination of two {\it mekal} and a power law model. The
best-fit parameters are listed in Table~\ref{t5}. For the best-fit
CMCD model, the inner disk temperature is $291_{-82}^{+93}\ev$.  The
disk inclination angle is poorly constrained.

\section{Individual ULX characteristics}
In this section we describe the X-ray properties of individual ULXs.
\subsection{NGC~5194 source 5}
This source is located in the western outer spiral arm of NGC~5194,
coincident with the \rosat{} HRI source 4 (Ehle et al. 1995).
\chandra{} resolved this source into two discrete sources NGC~5194\#5
(CXOM51~J132939.5+471244) and NGC~5194\#6 (CXOM51~J132940.0+471237).
NGC~5194\#5 is brighter than NGC~5194\#6 by factors of $1.5$ and $11$
in the $0.5-8\kev$ and $2-8\kev$ bands, respectively (Terashima \&
Wilson 2004).  Our \xmm{} spectrum of this source requires a soft
X-ray excess component that is equally well described either by an MCD
or a {\it mekal} component.  This soft excess component contributes
$\sim 42\%$ to the total flux in the $0.5-2\kev$. Therefore, the
detection of soft X-ray excess emission above the power-law component
in the \xmm{} spectrum could be due to the contamination of the softer
NGC~5194\#6 and the ULX spectrum is likely to be a simple power-law.
\subsection{NGC~5194 source 9}
This is the softest ULX in M~51 (see Fig.~\ref{f1}). This source is
coincident to the \rosat{} HRI source 5 (Ehle et al. 1995) and
\chandra{} source NGC~5194\#9 which was a super-soft source (Terashima
\& Wilson 2004). \chandra{} detected no photons above $1\kev$ from
NGC~5194 source 9 and its spectrum was well described by an MCD model
with $kT \sim 100\ev$. \xmm{} EPIC MOS detected this source at
$6\sigma$ and $3\sigma$ levels above $1\kev$ and $2\kev$,
respectively.  The MOS spectra of this source statistically require
three components: a power-law ($\Gamma \sim 1.4$), an MCD ($kT \sim
130\ev$), and a {\it mekal} ($kT \sim 340\ev$). Alternatively, the MOS
spectra can also be described by a power law ($\Gamma \sim 1.6$) and
two {\it mekal} components ($kT \sim 80$, $306\ev$). The X-ray flux of
source 5 was a factor of $\sim 3$ higher during our \xmm{} observation
than that during the \chandra{} observations. The soft X-ray excess
(MCD + {\it mekal}) flux is $7.4\times
10^{-14}{\rm~erg~cm^{-2}~s^{-1}}$ in the $0.5-8\kev$ band which is
similar to the total flux observed by \chandra{} in June 2001.

\subsection{NGC~5194 source 26}
This is the nearest ULX to the LLAGN in NGC~5194. It is located about
$28\arcsec$ west of the LLAGN and its soft X-ray emission is heavily
contaminated by extended soft emission. This source was not detected
in the \rosat{} HRI possibly due to the extended soft X-ray emission.
This ULX is brighter than the LLAGN in the $1.5-10\kev$ band (see
Fig.~\ref{f2}). \chandra{} also detected this source as the hardest
ULX in M~51 (Terashima \& Wilson 2004). In the two \chandra{}
observations, this source showed interesting spectral behavior. The
first \chandra{} observation revealed an absorbed power-law spectrum
and emission lines at $\sim 1.8\kev$, $3.24\kev$, $\sim4.03\kev$, and
$\sim6.65\kev$, most likely the K$\alpha$ lines of Si, Ar, Ca, and Fe,
respectively (Terashima \& Wilson 2004). In the second \chandra{}
observation, no emission lines were clearly detected although the
source was brighter by $\sim 50\%$. The \xmm{} spectra of this source
are well described by an absorbed power-law ($\Gamma \sim 2.4$, $N_H
\sim 7.1\times 10^{22}{\rm~cm^{-2}}$) and two {\it mekal} components
($kT \sim 250\ev$, $\sim 590\ev$). The absorbed power-law is
consistent with that observed by the second \chandra{} observation.
The two {\it mekal} components represent the contribution of the
extended soft emission.  We do not detect any emission lines except
the iron line or an edge.  Future deep X-ray observations will
establish if the line is real. The EPIC PN spectrum of source 26,
after appropriately correcting for the contribution of extended
emission, requires either a moderately broad ($\sigma \sim 200\ev$)
iron $K\alpha$ line or a strong iron K edge at $\sim 7.1\kev$.
   
\subsection{NGC~5194 source 37}
This source is coincident with a southern spiral arm of NGC~5194 that
displays extended UV emission (see Fig.~\ref{f2}). This source was not
detected in the first \chandra{} observation but was detected in the
second observation (Terashima \& Wilson 2004).  During our \xmm{}
observations, the ULX was $\sim 50\%$ brighter than during the second
\chandra{} observation.  The \chandra{} spectrum was described by a
simple power law, while the \xmm{} spectra statistically require a
soft component that is equally well described by a {\it mekal} or an
MCD component, in addition to a power law. The power law is flatter
($\Gamma = 0.8_{-0.5}^{+0.4}$) than that derived from the \chandra{}
data ($\Gamma = 1.55_{-0.15}^{+0.19}$). It is likely that this
discrepancy is attributable to the different number of spectral
components used to describe the \chandra{} and \xmm{} spectra.

\subsection{NGC~5194 source 41}
This ULX is closer to the nucleus of NGC~5195 than to the nucleus of
NGC~5194 but may be associated with the outer edge of an outermost
spiral arm to the north of the nucleus of NGC~5194. This source was
also detected in the \rosat{} HRI (source 10 in Ehle et al. 1995).
\chandra{} resolved this source into two sources NGC~5194\#41
(CXOM51~J132953.7+471436) and NGC~5194\#42 (CXOM51~J132953.8+471432).
NGC~5194\#41 is a factor of $\sim 2$ more luminous than NGC~5194\#42.
Remarkably, the \xmm{} source is fainter by $\sim 50\%$ than the
\chandra{} source NGC~5194\#41 alone. \chandra{} spectra of this ULX
are consistent with either a power-law or an MCD and no variability
between the two observations. \xmm{} spectra statistically require
either an MCD ($kT \sim 270\ev$) or a {\it mekal} ($kT \sim 340\ev$)
in addition to a power-law. Since the \chandra{} source NGC~5194\#42
is an absorbed source, the thermal component inferred from the \xmm{}
data is likely a genuine component.

\subsection{NGC~5194 source 63}
This is faintest ULX in M~51 with a $0.5-8\kev$ luminosity of $\sim
10^{39}{\rm~erg~s^{-1}}$. This source was fainter by a factor of $\sim
2$ in the $0.5-8\kev$ band during the \chandra{} observations and was
not considered to be a ULX. We consider this source to be a ULX at its
increased flux. It is possible that the source is a stellar mass black
hole accreting at or near the Eddington rate. The ULX is located to
the south-east region coincident with a spiral arm, displaying
extended UV emission. The X-ray spectrum of this source shows soft
excess emission above a power law ($\Gamma \sim 2.0$) and is
statistically better described by a {\it mekal} plasma than an MCD.
The presence of an extended UV source and optically thin thermal
plasma both suggest that this ULX may be associated with a region of
strong star-formation or unusually X-ray bright supernova remnant.

\subsection{NGC~5194 source 69}
This ULX is well known for the tentative detection of a $2.1{\rm~hr}$
periodicity in its X-ray flux (Liu et al. 2002). It is located in a
region of extended soft X-ray emission and strong UV emission in an
spiral arm to the north-east of the nucleus.
The ULX is closer to the nucleus of NGC~5195 than to the nucleus of
NGC~5194 but is clearly associated with NGC~5194.  This source was
detected in the \rosat{} HRI (Ehle et al. 1995).  The ULX showed
drastic long term X-ray variability, its $0.5-8\kev$ luminosity
declining by a factor of $\sim 75$ from the first to the second
\chandra{} observation. During our \xmm{} observation, the ULX had
brightened by a factor of $\sim 50$ in the $0.5-8\kev$ band. Source 69
also shows variability on short-times scales. The first \chandra{}
observation tentatively detected two peaks separated by $\sim 7000\s$.
Liu et al. (2002) interpreted these features as a $2.1{\rm~hr}$
periodicity. The number of counts was too low to detect any such
periodicity during the second \chandra{} observation. During a bright
flux state during the \xmm{} observation, the source again shows three
possible peaks separated this time by $\sim 6000\s$. We detect a
periodic signal at a significance level of $2\sigma$ with a period of
$5925\pm200\s$ from the EPIC light curves (see section~\ref{period}
for further discussion). The \xmm{} spectrum of this source is well
described by either of the following three component models: ($i$) a
power law ($\Gamma \sim 1.2$) and two {\it mekal} components ($kT \sim
180\ev$, $680\ev$); ($ii$) a power law ($\Gamma \sim 1.2$), an MCD
($kT \sim 170\ev$), and a {\it mekal} ($kT \sim 690\ev$).

\subsection{NGC~5194 source 82}
This is the brightest ULX in M~51 with an observed luminosity of $\sim
2.2\times 10^{39}{\rm~erg~s^{-1}}$ in the $0.5-8\kev$ band. The ULX is
coincident with an outer spiral arm east of the nucleus. This source
was previously detected with Einstein HRI (Palumbo et al. 1985),
\rosat{} HRI (source 14 in Ehle et al. 1995), \sax{} (Fukazawa et al.
2001), and \chandra{} (Terashima \& Wilson 2004). The source had
dimmed in the second \chandra{} observation but was brighter by $35\%$
in the \xmm{} observation, a level similar to that seen during the
first \chandra{} observation.  Both \chandra{} and \xmm{} spectra of
this ULX are consistent with an absorbed power law.

\subsection{NGC~5195 source 12}
This is the only ULX in NGC~5195 that is well separated from the
nucleus. The ULX is $\sim 70\arcsec$ away to east of the nucleus. The
ULX was earlier detected with the \rosat{} HRI (source 13 in Ehle et
al. 1995).
Source 12 is the only ULX in M~51 to have X-ray spectrum that is well
described by a single spectral component either a simple power-law
($\Gamma \sim 1.6\pm 0.2$) or {\it mekal} plasma ($kT \sim 7.7\kev$).
The ULX is not resolved spatially with \chandra{}.  Modeling the
spectrum as an optically thin plasma requires an unrealistically high
temperature and, as a result, we prefer the non-thermal power-law
interpretation despite the equally good spectral fits. The notable
lack of an optical/UV counterpart for the source strongly suggests
that this source is not a background AGN, particularly given its low
intrinsic X-ray absorption ($N_H \sim 5.6\times 10^{20}{\rm~cm^{-2}}$)

\section{Discussion \label{discuss}}
We have performed the temporal and spectral analysis of the LLAGN and
nine ULXs in M~51 using a \xmm{} observation.  Below we discuss the
main results obtained.

\subsection{The LLAGN of NGC~5194}
The X-ray spectrum of the LLAGN is a typical of Thomson-thick Seyfert
2 galaxies, namely an extremely flat continuum, a strong and narrow
line from neutral iron, and unabsorbed soft X-ray emission.  Our
spectral modeling shows that the flat continuum can be described as
the reflection of the primary power-law ($\Gamma = 1.9$) by cold and
dense material, assuming the cold material subtends a $2\pi$ solid
angle to the central source, the luminosity of the primary continuum
is then estimated to be $\sim 10^{40}{\rm~erg~s^{-1}}$ in the
$0.3-10\kev$ band.  This luminosity is about two orders of magnitude
lower than that for Seyfert 1 galaxies, thus confirming the nucleus of
NGC~5194 as an LLAGN as already established with previous \sax{} and
optical observations.  Since the LLAGN is nearly Thomson thick, the
soft X-ray emission in the spectrum must be associated with the
extended soft emission seen in Fig.~\ref{f1}.

\subsection{ULXs in M~51} 
There are nine ultra-luminous X-ray sources in M~51 with the observed
luminosity of $\gtsim 10^{39}{\rm~erg~s^{-1}}$ in the $0.5-8\kev$
band. Three ULXs, source 9, 41 and 63, with $L_X \sim
10^{39}{\rm~erg~s^{-1}}$ ($0.5-8\kev$) may actually be stellar mass
black holes ($\sim 10 - 20{\rm M_\odot}$). The brightest ULX in M~51,
source 82, has a $0.5-8\kev$ luminosity of $2.2\times
10^{39}{\rm~erg~s^{-1}}$, lower than the Eddington luminosity ($\simeq
2.6\times 10^{39}{\rm~erg~s^{-1}}$) for a $20{\rm~M\odot}$ black hole.
Thus, all ULXs in M~51 could plausibly be stellar mass black holes at
around (or slightly above) the Eddington limit.

The larger collecting area of \xmm{} compared to \chandra{} has
enabled us to investigate the nature of ULXs in greater detail. While
\chandra{} spectrum of individual ULXs in M~51 is well described by a
single spectral component, \xmm{} spectrum of most ULXs in NGC~5194
require at least two spectral components.
\subsubsection{Cool accretion disks}
X-ray spectra of two ULXs, sources 9 and 69, require soft components,
modeled by either {\it mekal} {\it plus} MCD or two {\it mekal}
components.  The MCD components suggest inner disk temperatures of
$kT_{in} \sim 130\ev$ (source 9) and $\sim 170\ev$ (source 69).  These
temperatures are lower than that found for ULXs studied with \asca{}
(Makishima et al. 2000) and Galactic X-ray binaries, but are similar
to that found for ULXs studied with \xmm{} (Miller et al. 2003;
Miller, Fabian, \& Miller 2004; Dewangan et al.  2004). The \xmm{}
observation of M~51 supports the increasing trend that ULXs have
cooler accretion disks than GBH binaries. The temperature of the
innermost regions of an accretion disk is given by
\begin{equation}
  kT_{\rm in}\approx 1.2\,{\rm keV}\left(\xi\over 0.41\right)^{1/2}
  \left(\kappa\over 1.7\right)\alpha^{-1/2}\left({\dot M}\over{{\dot M}_E}
  \right)^{1/4}\left({{M}\over{10\,M_\odot}}
  \right)^{-{{1}\over{4}}}
  \label{eq1}
\end{equation}
(see e.g., Makishima et al. 2000). Here $\kappa \simeq 1.7$ is a
spectral hardening factor. The factor $\xi \simeq 0.4$ takes into
account the fact that the radius of maximum temperature is larger than
the innermost stable orbit, and $\alpha = 1$ for a Schwarzschild
geometry or $1/6$ for a Kerr geometry.  Thus objects with lower MCD
temperatures must have either higher black hole masses or lower
accretion rates. Assuming a relative accretion rate of
$\frac{\dot{M}}{\dot{M}_{Edd}}\sim 0.1$, equation~\ref{eq1} implies
black hole masses of $\sim 7300 M\sun$ (source 9) and $\sim 2500
M\sun$ (source 69).

The results of the CMCD model fit to the spectrum of source 26
suggests that this ULX has cool accretion disk ($kT
=291_{-82}^{+13}\ev$).  Using the normalization and the inclination
angle of CMCD model, it is possible to derive a BH mass as detailed in
Wang et al. (2004). Using this procedure, we derive a black hole mass
of $\sim 2800 M\sun$ for the ULX sources 26.

\subsubsection{ULXs with a power-law spectrum}
The X-ray spectrum of two ULXs, source 82 and source 12, can be
described by a simple power-law models with a photon indices of $\sim
2.4$ and $\sim 1.7$, respectively.  Source 12 does not show any
signature of emission from an accretion disk. The power-law spectral
form of this ULX is similar to that of XRBs in their low state
(Terashoma \& Wilson 2004) or type 1 AGNs (excluding the narrow-line
Seyfert~1 galaxies). If the above analogy is correct, then the ULXs
with simple power-law spectrum may imply very high mass ($>$ a few
times $100M\sun$) black hole.  A power-law spectrum may also result
from beamed X-ray emission. The X-ray spectrum of source 82 is very
steep and may be similar to that of Galactic XRBs in their very high
state (see Terashima \& Wilson 2004 and Roberts et al. 2004 for a
detailed discussion on the power-law spectral form of ULXs).  It is
also possible that ULXs with a simple power-law spectrum are
background AGNs.

\subsubsection{Spectral variability}
\begin{figure}
  \centering \includegraphics[width=15cm]{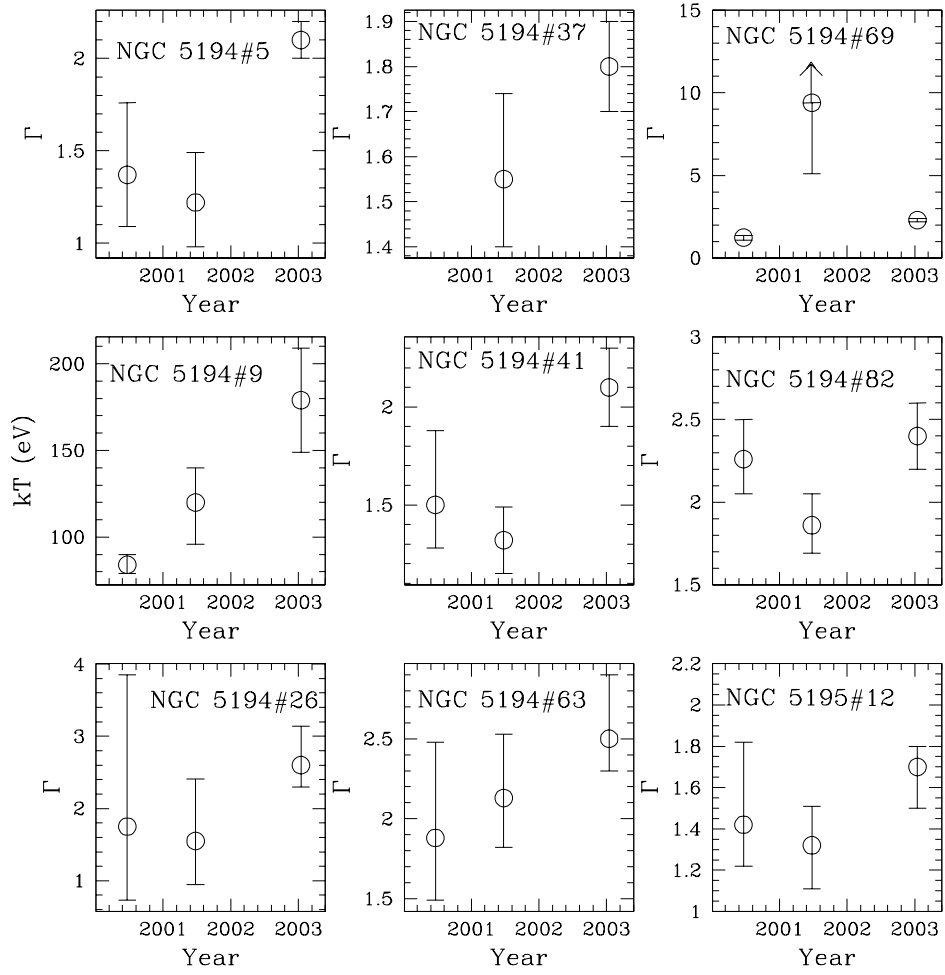}
\caption{Long-term spectral shape variability of ULXs in M~51. The best-fit
  power-law models (for all ULXs but NGC~5194 source 9) and MCD model (for
  NGC~5194 source 9) to the \chandra{} data of June 2000 and June 2001, taken
  from Terashima \& Wilson (2004), and
  to the \xmm{} data of January 2003 have been used.
  \label{f11}}
\end{figure}

It is not possible to compare directly the best-fit spectral form
derived from the \chandra{} and \xmm{} observations due to the
different number of spectral components used to fit the spectra.
Therefore, we compare the best-fit model derived from the \chandra{},
usually a single component model, and the corresponding single
component model best-fit to the \xmm{} data. The single component
model is not necessarily the best-fit model to the \xmm{} data.
Figure~\ref{f11} compares the spectral parameters illustrating the
spectral variability of ULXs in M~51. The steeper spectrum of source 5
during the \xmm{} observation may be due to the contribution of the
soft source NGC~5194\#6 detected with \chandra{} (see section 6.1).
Source 9 shows a very soft X-ray spectrum and is consistent with no
spectral shape variability. The detection of the power law component
in addition to the soft component in the spectra of source 9 may be
due to the larger collecting area of \xmm{} particularly at higher
energies.  Sources 26, 37, 63 and 12 do not show significant
variability in their spectral shape. Sources 41 and 82 show slightly
steeper X-ray spectra compared with the earlier \chandra{}
observation.

Source 69 continues to show drastic spectral and flux variability. The
source was in a high/hard state (power law $\Gamma =
1.24_{-0.17}^{+0.12}$ or MCD $kT = 2.3_{-0.5}^{+1.0}$) in June 2000
and subsequently made a transition to an extremely soft/low state
(power law $\Gamma > 5.1$ or MCD $kT = 0.17_{-0.06}^{+0.13}$) in June
2001 (Terashima \& Wilson 2004). The ULX made another transition back
to a state similar to its earlier high/hard state, in January 2003.
The spectral transitions of source 69 are completely different from
the high/soft to low/hard spectral transitions usually observed from
Galactic XRBs but similar to those observed from some ULXs (The
Antennae, Fabbiano et al. 2003; Holmberg II X-1, Dewangan et al. 2004;
an ULX in NGC~7714, Soria \& Motch 2004). The ULXs that show distinct
spectral transitions may a comprise a distinct class of ULXs. The
physics of these transitions is yet to be understood clearly.
   
\subsubsection{Periodicity \label{period}}
Source 69 is the only ULX in M~51 that shows indications for a
possible periodic variations in its X-ray flux.  The periods detected
from our \xmm{} observation ($P=5925\pm200\s$ at a $\sim 2\sigma$
level) and the first \chandra{} observation ($P=7620\pm500\s$) are
significantly different. The two observations are separated by $\sim
1.5{\rm~yrs}$.  The variation in period, if real, clearly argues
against orbital periodicity, and suggests the presence of
quasi-periodic oscillations.  Liu et al. (2002) discussed the
possibility that the observed periodicity is due to orbital variations
and as noted by Liu et al. (2002), only low mass X-ray binaries
(LMXBs) can have orbital periods as small as $2{\rm~hr}$ with high
luminosities.  In this case Liu et al. (2002) estimate the mass of the
donor to be $\simeq 0.23M\sun$.  The spectrum of source 69 derived
from the first \chandra{} observation was found to be consistent with
an absorbed multicolor disk model with inner disk temperature of $kT =
2.10_{-0.40}^{+0.63}\kev$ and showed marginal improvement in the fit
after including a single temperature blackbody component. Based on
these results, Liu et al. (2002) do not rule out the possibility that
source 69 is a neutron star. Our \xmm{} spectra require three
components - a power-law ($\Gamma \sim 1.2$), a {\it mekal} ($kT\sim
680\ev$) and another {\it mekal} ($kT \sim 180\ev$) or a cool
accretion disk ($kT_{in} \sim 170\ev$). The stellar mass black hole
systems and high luminosity neutron star binary systems, both have
high disk temperatures ($kT = 1-2\kev$).  The possible variation in
the period, due to a QPO and the X-ray spectrum both are inconsistent
with the LMXB picture. Moreover, source 69 resides in a region that is
likely to be undergoing active star formation, making it likely that
the ULX is a high mass X-ray binary (HMXB). The spectral transition of
source 69 from a high-hard to low-soft state also is in disagreement
with the behavior of most Galactic LMXBs (Terashima \& Wilson 2004).

If we interpret the variation of the apparent periodicity as resulting
from QPOs, then QPO frequency of $\sim 0.15 mHz$ is much lower than
$67mHz$ QPO observed with \rxte{} from the Galactic black hole
candidate GRS~1915+105, which has a dynamically measured mass of
$14\pm4 M\sun$ (Morgan, Remillard \& Grener 1997).  The only ULX known
to show QPO is M~82 X-1 (Strohmayer \& Mushotzky 2003). Its QPO
frequency of $54mHz$ is also lower than that of GRS~1915+105 but much
larger than the implied frequency for source 69. It is not necessary
the case that QPO frequency scales with the mass of the black hole,
only twin peak QPOs with a frequency ratio of 3:2 have been
established to scale with black hole mass (McClintock \& Remillard
2003). The micro-quasar GRS~1915+105 is known to show QPOs at kHz as
well as mHz frequencies, the lowest frequency of QPO observed from GRS
1915+105 is $1.6{\rm~mHz}$ (Morgan et al. 1997).  Therefore it is not
entirely unlikely the possibility of a $\sim 0.15{\rm~mHz}$ QPO from
source 69. Long observations are required to clarify the presence of
periodicity or QPO in this ULX.

\section{Conclusions \label{conclusion}}
We have analyzed \xmm{} EPIC observation of M~51 performed in January
2003. Our main findings are as follows:
\begin{itemize}
\item We detected 9 ULXs, an LLAGN and extended soft thermal emission
  from M~51. One of the sources (source 63) is a new ULX, not
  identified earlier due to its previously faint flux levels.

\item All eight ULXs in NGC~5194 are located in or near spiral arms.
  Four ULXs are located in the regions of strong UV emission in the
  spiral arms, suggesting association of these ULXs with current star
  formation.

\item Two ULXs (sources 26 and 69) show evidence for short-term
  variability. Source 69 is the most rapidly variable ULX in M~51 and
  shows a possible period of $\sim 5925\s$. All the ULXs except
  NGC~5194\#12, show long-term variability by a factors of few to
  several.

\item Most ULXs in M~51 show soft X-ray excess emission, above a hard
  power law, that is modeled by a {\it mekal} plasma and/or an MCD
  component. However the strength of the soft excess emission inferred
  from an ULX is comparable to the diffuse emission around individual
  ULXs excluding source 69. Thus the soft excess emission is unlikely
  to be physically associated with any of the ULXs except source 69.
  The soft excess emission from source 69 is about a factor four
  stronger than the surrounding diffuse emission and is likely
  intrinsic to the ULX.

\item There is an indication of either a moderately broad ($\sigma =
  100\ev$) iron K$\alpha$ line or an iron K absorption edge both due
  to neutral iron in the spectrum of source 26.
\item The X-ray spectrum of LLAGN in NGC~5194 is extremely flat
  ($\Gamma \sim 0.7$) and shows a strong (EW$\sim 2.5\kev$), narrow,
  iron line from neutral material.  The X-ray emission from the LLAGN
  in NGC~5194 is dominated by reflection from cold material.
\end{itemize}

\acknowledgements We are grateful to an anonymous referee for the
valuable comments and suggestions that improved the paper. This work
is based on observations obtained with \xmm{}, an ESA science mission
with instruments and contributions directly funded by ESA member
states and the USA (NASA).  GCD acknowledges the support of NASA grant
through the award NNG04GN69G. REG acknowledges NASA award NAG5-9902 in
support of his Mission Scientist position on \xmm{}. This research has
made use of data obtained from the High Energy Astrophysics Science
Archive Research Center (HEASARC), provided by NASA's Goddard Space
Flight Center.

\clearpage

\clearpage

\begin{table}
  \centering
  \caption{The low luminosity AGN and list of ULXs in M~51 \label{t1}}
{\footnotesize
  \begin{tabular}{lcccccccc}
    \tableline\tableline
    \chandra{}  & Host    & \xmm{} & RA (2000) & Decl. (2000) & \multicolumn{3}{c}{$2-10\kev$ count rate} \\
         id     & galaxy  &  name           &           &              &  \multicolumn{3}{c}{($10^{-2}{\rm~counts~s^{-1}}$)} \\       
                &         &  XMMU           &           &              &   PN    &    MOS1    &    MOS2        \\ \tableline  
    LLAGN & NGC~5194  & J132952.9+471140  & 13:29:52.87 & +47:11:40.09 & $1.43\pm0.09$ & $0.46\pm0.05$ &  $0.46\pm0.05$ \\
    5  & NGC~5194  & J132939.5+471241   & 13:29:39.55 & +47:12:41.09  & $0.97\pm0.09$ & $0.38\pm0.05$ & $0.27\pm0.04$  \\
    9\tablenotemark{a}  &  NGC~5194 &  J132943.4+471134  & 13:29:43.37 & +47:11:34.21 &  --   &  $0.08\pm0.03$  &  $0.05\pm0.02$  \\
    26 &  NGC~5194 &  J132950.7+471153  & 13:29:50.73 &  +47:11:52.86  & $2.40\pm0.11$ &  $0.78\pm0.06$ & $0.82\pm0.06$ \\
    37\tablenotemark{a} &   NGC~5194 & J132953.3+471040 &  13:29:53.26 & +47:10:40.23    & --    &  $1.64\pm0.09$ & $1.64\pm0.09$   \\
    41 &   NGC~5194 & J132953.8+471433 & 13:29:53.77 &  +47:14:33.01    &  $0.71\pm0.06$ & $0.17\pm0.03$ & $0.19\pm0.03$  \\
    63 &   NGC~5194 & J132957.6+471047  &  13:29:57.56 & +47:10:46.62  & $0.47\pm0.05$ &  $0.18\pm0.03$ & $0.18\pm0.03$ \\
    69 &  NGC~5194 & J133001.1+471342   & 13:30:01.090 & +47:13:41.47  & $1.37\pm0.09$ & $0.40\pm0.05$ &  $0.33\pm0.04$ \\
    82\tablenotemark{a} &  NGC~5194 & J133007.4+471104  & 13:30:07.407 &  +47:11:03.94  &   --  & $0.50\pm0.05$ & $0.49\pm0.05$  \\
    12 &  NGC~5195 & J133006.1+471540  &  13:30:06.13 &  +47:15:40.20 & $0.83\pm0.07$ &  $0.25\pm0.04$ & $0.27\pm0.04$  \\
    \tableline
  \end{tabular}}
\tablenotetext{a}{These sources fall on the PN chip gaps. The coordinate was derived from the full band image. }
\end{table}

\begin{table}
  \centering
  \caption{Results of short-term variability tests. \label{t2}}
  \begin{tabular}{lccc}
    \tableline \tableline
    & \multicolumn{3}{c}{$\chi^2$ statistic}    \\
    Source & count rate & $\chi^2$/dof & $P_{\chi^2}$ (var) \\
    & (${\rm count~s^{-1}}$)     &   &        \\ \tableline 
    5 &  $0.063_{-0.002}^{+0.002}$ & 29.29/36  & --   \\
    9 & $0.044_{-0.001}^{+0.001}$  & 42.88/36  & --   \\
    26 & $0.070_{-0.002}^{+0.002}$ & 57.28/36  & $98.64\%$ \\
    37 & $0.053_{-0.002}^{+0.002}$ & 41.66/36  & --    \\
    41 & $0.023_{-0.001}^{+0.001}$ & 46.29/36 & --  \\
    63 & $0.034_{-0.001}^{+0.001}$ & 30.77/36 & --  \\
    69 & $0.049_{-0.002}^{+0.002}$ & $70.91/36$ & $99.95\%$ \\
    82 & $0.074_{-0.002}^{+0.002}$ & 42.23/36 & -- \\
    12 & $0.025_{-0.001}^{+0.001}$ & 38.19/36 & --  \\
    \tableline
  \end{tabular}
\end{table}

\begin{table}
  \centering
  \caption{Results of the joint spectral fittings to the EPIC PN and MOS spectra
    of M~51 nucleus \label{t3}}
{\scriptsize 
 \begin{tabular}{lccccc}
    \tableline\tableline
     Component & Parameter  & model 1\tablenotemark{a} & model 2\tablenotemark{a} & model 3\tablenotemark{a} \\
    \tableline
    absorption & $N_H{\rm~cm^{-2}}$ & $1.57\times 10^{20}$(f) &  $1.57\times10^{20}$(f) & $1.57\times 10^{20}$(f) \\
    {\it mekal} & $kT(\ev)$ &  $171_{-29}^{+23}$ & $603_{-17}^{+16}$   & $177_{-32}^{+36}$    \\
    & Abundance (solar) & $0.20_{-0.14}^{+2.49}$  & --   &  $0.20(f)$ \\
    & O  & -- &  $0.08_{-0.06}^{+0.07}$ & -- \\
    & Ne & -- &  $0.18_{-0.07}^{+0.07}$ & -- \\
    & Mg  & -- &  $0.10_{-0.05}^{+0.05}$ & -- \\
    & Si  & -- &  $0.12_{-0.05}^{+0.06}$ & -- \\
    & S  & --  & $0.32_{-0.22}^{+0.29}$ & -- \\
    & Ar  & -- & $1.47_{-1.08}^{+1.28}$  & -- \\
    & Ca  & -- & $2.38_{-1.07}^{+3.14}$ & -- \\
    & Fe & -- & $0.11_{-0.01}^{+0.02}$ & -- \\
    &  Ni  &  -- & $< 0.17$  &  -- \\
    & $norm$\tablenotemark{c} & $2.0_{-1.4}^{+4.2}\times10^{-4}$ & $1.2_{-0.1}^{+0.1}\times 10^{-3}$  & $1.9_{-0.3}^{+0.6}\times 10^{-4}$ \\
 {\it mekal} & $kT(\ev)$ & $610_{-14}^{+16}$ & --  & $614_{-11}^{+13}$  \\
    & Abundance & $0.18_{-0.02}^{+0.05}$ & -- & $0.18(f)$ \\
    & $norm$\tablenotemark{c} & $7.6_{-1.3}^{+1.7}\times 10^{-4}$ & --  & $7.8_{-0.3}^{+0.3}\times 10^{-4}$\\ 
  Power law & $\Gamma$ & $0.43_{-0.33}^{+0.34}$   & $0.2_{-0.4}^{+0.3}$ & --  \\
    & norm & $4.1_{-1.7}^{+1.5}\times 10^{-6}$  & $2.7_{-1.3}^{+1.9}\times 10^{-6}$ & -- \\
    pexrav & $\Gamma$ & --  &  -- &  $1.9(f)$  \\
    & $R$  & --  & --  &  $1.0(f)$ \\
    & $i$(${\rm degree}$)  &   --   &  --      &  $70(f)$  \\ 
    & norm & --   &  --        & $2.1_{-0.2}^{+0.3}\times10^{-4}$\\
    Gaussian &  $E(\kev)$  &  $6.43_{-0.03}^{+0.03}$   &   $6.43_{-0.03}^{+0.02}$ & $6.43_{-0.03}^{+0.02}$   \\
    & $\sigma$ ($\ev$)  & $61_{-51}^{+42}$  & $62_{-50}^{+42}$  &  $10(f)$  \\
    & $f_{line}\tablenotemark{d}$    &  $4.7_{-1.0}^{+1.1}$  & $4.74_{-1.00}^{+1.14}$    & $4.6_{-1.1}^{+0.9}$ \\
    & EW ($\kev$) & $2.6$  & $2.6$    & $2.5$ \\
    Total & $f(0.3-10\kev)\tablenotemark{e}$ & $8.2$  & $8.2$   & $8.2$  \\
          & $L(0.3-10\kev)\tablenotemark{f}$ & $6.9$  & $6.9$   & $6.9$  \\
    & $f(2-10\kev)\tablenotemark{e}$  & $2.1$ & $1.9$  &  $2.0$ \\
    & $L(2-10\kev)\tablenotemark{f}$  & $1.8$ & $1.6$  &  $1.7$  \\ 
    & $C/dof$ & $414.7/401$  & $414.7/396$  &  $420.9/400$  \\
    & Goodness of fit & $47.8\%$ & $46.9\%$ & $61.8\%$ \\   \tableline
  \end{tabular}}
\tablenotetext{a}{(f) indicates a fixed parameter.}
\tablenotetext{b}{Model 1 is a combination of two {\it mekal}, a power-law, a narrow
  Gaussian multiplied the Galactic absorption
  model. Model 2 is the same as model 1 except that the two {\it mekal} components have been replaced by a {\it vmekal} component. Model 3 is the same as model 1 with the power-law 
  replaced by a neutral reflection model {\it pexrav}. }
\tablenotetext{c}{In units of $10^{-14}\frac{1}{4\pi
    d^2}\int {n_e n_H dV}$, where $d$ is the angular size distance to the
  source (${\rm~cm}$), $n_e$ is the electron density (${\rm~cm^{-3}}$), and
  $n_H$ is the hydrogen density (${\rm~cm^{-3}}$). }
\tablenotetext{d}{Line flux in units of $10^{-6}{\rm~photons~cm^{-2}~s^{-1}}$. }
\tablenotetext{e}{Observed flux in units of $10^{-13}{\rm~erg~cm^{-2}~s^{-1}}$}
\tablenotetext{f}{Observed luminosity in units of $10^{39}{\rm~erg~s^{-1}}$.}
\end{table}

\clearpage
\begin{deluxetable}{llllllllllll}
  \tabletypesize{\footnotesize}%
\tablecolumns{12} 
\tablewidth{0pc} 
\tablecaption{Best-fit spectral model parameters derived from the joint PN/MOS data for the ULXs in M~51. The fluxes measured with the PN and MOS data are similar within errors, and the quoted fluxes are from the PN data only. \label{t4}}
\tablehead{
\colhead{ULX} &  \colhead{Model\tablenotemark{a}} & \colhead{N$_{\rm H}$\tablenotemark{b}} & \colhead{$\Gamma$}  & \colhead{$kT_{MCD}$} & \colhead{$kT_{\it mekal}$} &  \colhead{$f_{X}$\tablenotemark{d}} & \colhead{$L_{X}$\tablenotemark{e}} & \colhead{$C/dof$} & \colhead{GOF} & \colhead{$\Delta C/\Delta p$\tablenotemark{f}} & \colhead{MLR}  \\
\colhead{} & \colhead{}  & \colhead{$10^{20}{\rm~cm^{-2}}$} & \colhead{}  & \colhead{($\ev$)} & \colhead{($\ev$)}  & \colhead{} &  \colhead{} &  \colhead{} & \colhead{} & \colhead{} & \colhead{significance} }   
\startdata

5  & A & $8.6_{-2.3}^{+2.6}$ & $2.5_{-0.1}^{+0.2}$ & -- & -- &  -- & --  & 234.6/206 & $76.4\%$ & --  & -- \\
& B & $1.57$(f) & --  &  $485_{-26}^{+28}$  &  -- &  --  & -- & 400.5/207  & $100.0\%$  & -- & -- \\
& C & $9.1_{-4.5}^{+4.4}$ & $2.0_{-0.1}^{+0.2}$  & $210_{-7}^{+45}$ & --  & $2.1_{-0.2}^{+0.3}$ & $1.8_{-0.2}^{+0.2}$   & 213.3/204  & $34.2\%$ & $-21.3/2$ & $2.4\times 10^{-5}$   \\
& D & $1.57$(f)  & $2.0_{-0.1}^{+0.1}$  & -- &  $802_{-111}^{+106}$ & $2.0_{-0.2}^{+0.2}$ & $1.7_{-0.2}^{+0.2}$  & 210.5/205 & $42.1\%$ & $-24.1/1$ & $9.1\times10^{-7}$  \\ \\

9\tablenotemark{c} & A & $28.6_{-10.0}^{+15.4}$  & $5.7_{-0.8}^{+1.3}$ & -- & --  & -- & -- & 113.7/54  & $99.8\%$ & -- & -- \\
& B  & $8.3_{-6.2}^{+8.4}$ & --  & $163_{-26}^{+29}$ & -- & -- & -- & 110.7/54 & $92.3\%$  & -- & --  \\
& C  & $28.4_{-13.8}^{+19.8}$  & $2.0_{-0.7}^{+0.8}$  & $107_{-25}^{+28}$ & -- & $1.0_{-0.5}^{+0.4}$ & $0.8_{-0.5}^{+0.3}$  &  69.5/52 & $72.9\%$ & $-44.2/2$ &  $2.5\times 10^{-10}$ \\
& D & $1.57(f)$ & $3.3_{-0.5}^{+0.5}$  & -- & $290_{-30}^{+62}$  & $0.9_{-0.2}^{+0.2}$ & $0.7_{-0.2}^{+0.2}$ & 72.9/53 & $89.3\%$ & $-40.8/1$ & $1.7\times10^{-10}$ \\
& E & $1.6_{-1.6}^{+1.8}$ & $1.4_{-0.8}^{+0.7}$ & $130_{-50}^{+36}$ & $349_{-42}^{+61}$ & $0.9_{-0.2}^{+0.2}$ & $0.8_{-0.2}^{+0.2}$ & 59.8/50 & $46.9\%$ & $-53.9/4$ & $5.5\times10^{-11}$ \\
& F&  $1.57$(f) & $1.6_{-0.8}^{+1.0}$ & -- & $80_{-44}^{+35}$,$306_{-39}^{+62}$ & $1.0_{-0.2}^{+0.3}$ & $0.8_{-0.2}^{+0.2}$ & 67.9/51 & $78.2\%$ & $-45.8/3$ & $6.2\times10^{-10}$ \\
\\
37\tablenotemark{c} &  A  &  $1.57$(f)  & $1.7_{-0.1}^{+0.1}$  &  -- & -- &  -- & -- & 104.9/61 & $99.9\%$ & -- & -- \\
& B & $1.57$(f) & -- & $1253_{-131}^{+156}$ & -- & -- & -- & 232.8/61 & $100.0\%$ & -- & -- \\
& C & $12.4_{-11.0}^{+15.9}$ & $0.8_{-0.5}^{+0.4}$ & $220_{-67}^{+125}$  & --  &  $2.3_{-1.1}^{+0.4}$ & $1.9_{-0.9}^{+0.3}$   & 68.6/58 & $57.9\%$ & $-36.3/3$ & $6.5\times10^{-8}$  \\
& D &  $1.57$(f)  & $0.8_{-0.5}^{+0.3}$ & --  & $691_{-94}^{+170}$  & $2.3_{-1.5}^{+0.5}$  & $1.9_{-1.2}^{+0.5}$  &  65.2/59 &  $65.2$ & $-39.7/2$ & $2.4\times 10^{-9}$  \\
\\
41    & A & $1.57$(f)           &  $1.8_{-0.1}^{+0.1}$ & --               & -- &  --              & -- &  169.1/129  &  $96.3\%$ & -- & -- \\
& B & $1.57$(f)           & --                 & $1134_{-82}^{+94}$ & -- & --               & -- & 392.7/129   & $100.0\%$ & -- & --  \\
& C & $4.2_{-4.2}^{+11.2}$  & $1.2_{-0.3}^{+0.3}$ & $272_{-100}^{+88}$ & -- & $1.1_{-0.5}^{+0.2}$ & $1.0_{-0.4}^{+0.2}$  & 143.8/126  & $59.8\%$ & $-25.3/3$ & $1.3\times 10^{-5}$  \\
& D  &  $1.57$(f)  & $1.5_{-0.2}^{+0.1}$ & --  & $341_{-73}^{+86}$  & $1.1_{-0.2}^{+0.1}$ & $0.9_{-0.2}^{+0.1}$ & 146.6/127 & $69.6\%$ & $-22.5/2$ & $1.3\times 10^{-5}$ \\
\\

63 & A  &  $5.4_{-2.7}^{+2.8}$ & $2.3_{-0.2}^{+0.1}$ & -- & --  & -- & --  & 137.0/101 &  $97.0\%$ & -- & -- \\
&  B & $1.57$(f) & --  &   $788_{-47}^{+55}$ & -- & --     & -- & 376.8/102  & $100.0\%$ & -- & --  \\
& C & $3.9_{-3.9}^{+10.0}$ &  $1.5_{-0.4}^{+0.5}$  & $283_{-105}^{+100}$ & -- & $1.1_{0.5}^{+0.2}$ & $0.9_{-0.4}^{+0.2}$ & 116.3/99 & $65.7\%$ & $-20.3/2$ & $3.9\times 10^{-5}$   \\
& D &  $1.57(f)$ & $1.8_{-0.2}^{+0.2}$  & -- & $608_{-149}^{+28}$    &  $1.1_{-0.2}^{+0.2}$   & $0.9_{-0.2}^{+0.1}$ &  105.1/100  & $42.60\%$ & $-31.9/1$ & $1.6\times 10^{-8}$ \\

69 &  A  & $1.57$(f) &  $2.1_{-0.1}^{+0.1}$ & -- & -- &  -- & -- & 342.0/204 & $100.0\%$  & -- & -- \\
& B & $1.57$(f)  & -- & $891_{-39}^{+42}$ & --  & -- & -- & 1091.3/204 &  $100.0\%$ & -- & --   \\
& C & $12.6_{-4.5}^{+7.0}$  & $1.3_{-0.2}^{+0.2}$ & $190_{-31}^{+28}$ & -- & $2.3_{-0.8}^{+0.3}$  &  $1.9_{-0.7}^{+0.2}$ &  $220.0/201$ & $67.3\%$ & $-122.0/3$ & $2.8\times 10^{-26}$ \\
& D & $1.57$(f) & $1.6_{-0.1}^{+0.1}$  & -- & $368_{-52}^{+44}$ & $2.2_{-0.3}^{+0.2}$  & $1.8_{-0.2}^{+0.2}$ & $244.6/202$ & $87.6\%$ & $-106.4/2$ & $7.8\times 10^{-24}$  \\
& E & $5.2_{-5.2}^{+8.8}$ & $1.2_{-0.2}^{+0.2}$ & $169_{-44}^{+73}$ & $689_{-73}^{+105}$ & $2.2_{-0.6}^{+0.3}$ & $1.9_{-0.5}^{+0.2}$ & 200.9/199 & $29.2\%$ & $-141.1/5$ & $1.0\times10^{-28}$ \\
& F & $1.57$(f) & $1.2_{-0.2}^{+0.2}$ & -- & $181_{-47}^{+57}$, $676_{-60}^{+112}$ & $2.3_{-0.4}^{+0.3}$ & $1.9_{-0.4}^{+0.2}$ & 191.9/200 & $17.1\%$ & $-150.1/4$ & $1.9\times10^{-31}$  \\
\\
82\tablenotemark{c} & A  & $15.8_{-3.7}^{+3.7}$  & $2.4_{-0.2}^{+0.2}$ & -- & --  & $2.6_{-0.4}^{+0.4}$ & $2.2_{-0.4}^{+0.4}$ &  $98.3/100$ &  $32.4\%$ & -- & -- \\
&  B & $1.57$(f) & -- & $809_{-56}^{+62}$  & -- & -- & -- & $164.2.3/101$ & $100.0\%$ & -- & -- \\
& C & $7.4_{-5.2}^{+8.2}$ & $1.4_{-0.6}^{+0.6}$ & $406_{-20}^{+43}$ & --  & $2.7_{-2.0}^{+0.1}$ & $2.3_{-1.7}^{+0.1}$  & $91.5/98$ & $12.8\%$ & $-6.8/2$ & $0.033$ \\
& D & $13.0_{-4.3}^{+4.4}$ & $2.2_{-0.2}^{+0.1}$ & --  & $727_{-258}^{+360}$ & $2.6_{-0.6}^{+0.4}$ & $2.2_{-0.5}^{+0.3}$  & 93.7/98  & $21.3\%$  & $-4.6/2$ & $0.100$  \\
\\
\tableline
12 & A & $5.6_{-3.5}^{+3.7}$ & $1.6_{-0.1}^{+0.1}$ & --  & -- & $1.2_{-0.3}^{+0.2}$ & $1.0_{-0.3}^{+0.2}$ & 88.4/93 & $21.3\%$ & -- & --  \\
& B & $1.57$(f) & -- & $1290_{-120}^{+140}$ &  -- & -- & -- & 121.5/98 & $92.2\%$  & -- & --  \\
& G & $1.57$(f) & -- &  -- & $7930_{-1637}^{+2923}$ & $1.3_{-0.3}^{+0.2}$ &  $1.1_{-0.2}^{+0.2}$ & 87.7/94  & $18.7\%$ &  -- & --   \\
\tableline \enddata
\tablenotetext{a}{A= PL, B = MCD, C = PL+MCD, D = PL+{\it mekal}, E =
  PL+MCD+{\it mekal}, F=PL+{\it mekal} + {\it mekal}, G={\it mekal}.}
\tablenotetext{b}{Total absorption column density. The Galactic column
  is $1.5\times10^{20}{\rm~cm^{-2}}$.}  \tablenotetext{c}{Only MOS
  data were used for the spectral fitting. The source photons fall at
  the chip gap in the PN camera.}  \tablenotetext{d}{Observed flux in
  units of $10^{-13}{\rm~erg~cm^{-2}~s^{-1}}$ and in the energy band
  of $0.5-8\kev$ band.}  \tablenotetext{e}{Observed luminosity in
  units of $10^{39}{\rm~erg~s^{-1}}$ and in the energy band of
  $0.5-8\kev$ band. \tablenotetext{f}{$\Delta C/\Delta p$ is in
    comparison to the PL (model A).}}
\end{deluxetable}

\begin{table}
  \centering
  \caption{Results of the joint spectral fittings to the EPIC PN and MOS spectra
    of the ULX NGC~5194 source 26 \label{t5}}
  \begin{tabular}{lcccc}
    \tableline\tableline
    Component & Parameter\tablenotemark{a} & model 1\tablenotemark{b} & model 2\tablenotemark{b} \\ 
    \tableline\tableline
    {\it mekal} & $kT(\ev)$ &  $253_{-69}^{+84}$  & $251_{-69}^{+86}$ \\
    & abundance (solar) & $0.1(f)$ & $0.1(f)$ \\
    & $norm$\tablenotemark{c} & $1.8_{-0.6}^{+1.2}\times10^{-4}$  & $1.8_{-0.6}^{+1.3}\times 10^{-4}$  \\
    {\it mekal} & $kT(\ev)$ & $589_{-40}^{+74}$ & $589_{-39}^{+75}$ \\
    & abundance (solar) & $0.1(f)$ & $0.1$ \\
    & $norm$\tablenotemark{c} & $3.4_{-1.2}^{+0.5}\times 10^{-4}$  & $3.4_{-1.2}^{+0.5}\times 10^{-4}$  \\ 
    absorption &  $N_H$ ($\times 10^{22}{\rm~cm^{-2}}$) &  $7.1_{-1.4}^{+1.5}$ & $10.5_{-4.2}^{+1.4}$ \\
    power law & $\Gamma$ &  $2.45_{-0.20}^{+0.39}$ & -- \\
    & $norm$ & $3.9_{-1.8}^{+3.7}\times 10^{-4}$  &  -- \\
    Gaussian &  $E\kev$  & $6.33_{-0.13}^{+0.11}$ & $6.33_{-0.12}^{+0.11}$ \\
    & $\sigma$ ($\ev$)  & $158_{-155}^{+180}$   & $170_{-136}^{+167}$ \\
    & $f$  & $2.3_{-1.2}^{+1.5}\times10^{-6}$ & $2.7_{-1.3}^{+1.5}\times 10^{-6}$ \\
    & EW ($\ev$) & $550$  & $615$  \\
    CMCD     & $kT_{in}$ ($\ev$) &  -- & $291_{-82}^{+13}$ \\
    & $kT_c$ ($\kev$) & -- & $50.0_{-18.1}^{+35.7}$ \\
    & $R_c$ ($R_g$) &  -- &  $6.65_{-3.2}^{+3.2}$  \\
    & $\tau$ &  -- &   $2.0_{-1.2}^{+3.0}$ \\
    & $i\tablenotemark{d}$  &  -- & $53.6_{-18.6p}^{+31.4p}$ \\
    & $K$\tablenotemark{e} &   -- & $143.9_{-90.3}^{+383.9}$ \\
    Total & $f_{X}(\times 10^{-13}{\rm~erg~cm^{-2}~s^{-1}})\tablenotemark{g}$ & $4.47_{-1.70}^{+0.65}$  & $4.47_{-2.34}^{+3.20}$ \\
    & $L_{X}(\times 10^{39}{\rm~erg~s^{-1}})\tablenotemark{g}$ & $3.76_{-1.42}^{+0.54}$ & $3.76_{-1.96}^{+2.69}$ \\
    & $C/dof$ &  278.5/295  & 275.0/291 \\ 
    & Goodness of fit & $9.3\%$ & $10.6\%$ \\ \tableline
  \end{tabular}
  \tablenotetext{a}{(f) indicates a fixed parameter.}
  \tablenotetext{b}{Model 1 is a combination of two {\it mekal}, a power-law, a narrow
    Gaussian multiplied by  an absorption
    model. Model 2 is same as model 1 except that the power-law has been replaced with the comptonized MCD model. }
  \tablenotetext{c}{In units of $10^{-4}\times10^{-14}\frac{1}{4\pi
      d^2}\int {n_e n_H dV}$, where $d$ is the angular size distance to the
    source (${\rm~cm}$), $n_e$ is the electron density (${\rm~cm^{-3}}$), and
    $n_H$ is the hydrogen density (${\rm~cm^{-3}}$). }
  \tablenotetext{d}{Parameter pegged at both limits}.
  \tablenotetext{e}{CMCD normalization, $K = (\frac{R_{in}/km}{D/10{\rm~kpc}}$).} 
  \tablenotetext{g}{Observed flux  and luminosity in the  $0.5-8\kev$ band.}
\end{table}


\end{document}